# Tunable and Persistent Macroscopic Polarization in Nominally Centrosymmetric Defective Oxides

Dae-Sung Park,[1,2,3]* Nini Pryds,[1] Nicolas Gauquelin,[4] Mahmoud Hadad,[3] Dmitry Chezganov,[4] Alessandro Palliotto,[1] Daen Jannis,[4] Jorge Íñiguez-González,[5,6] Johan Verbeeck,[4] Paul Muralt,[3] Dragan Damjanovic[2]*

1. *Department of Energy Conversion and Storage, Technical University of Denmark, Fysikvej, 2800 Kongens Lyngby, Denmark.*
2. *Group for Ferroelectrics and Functional Oxides, Swiss Federal Institute of Technology–EPFL, 1015 Lausanne, Switzerland.*
3. *Group for Electroceramic Thin Films, Swiss Federal Institute of Technology–EPFL, 1015 Lausanne, Switzerland.*
4. *Electron Microscopy for Materials Science (EMAT), University of Antwerp, B-2020 Antwerpen, Belgium.*
5. *Luxembourg Institute of Science and Technology, 5 Avenue des Hauts-Fourneaux, L-4362 Esch/Alzette, Luxembourg.*
6. *Department of Physics and Materials Science, University of Luxembourg, 41 Rue du Brill, L-4422 Belvaux, Luxembourg.*

**E-mail: daepa@dtu.dk ; dragan.damjanovic@epfl.ch*

## Abstract

**Introducing symmetry breaking in materials enables the emergence of functionalities. This can be microscopically and macroscopically driven by applying external stimuli such as mechanical stress, electric field, temperature, and chemical modification. For instance, non-zero net dipole moments are formed in a material with the presence of local charged defects or their clusters, which can alter the crystal structure, charge states, and electrostatic potential across the material. Here, we demonstrate a conceptual approach to defects-mediated symmetry breaking that allows for built-in polarization in a nominally centrosymmetric defective oxide, $Gd_xCe_{1-x}O_{2-\delta}$ (CGO) films via creating a macroscopic charge asymmetry. Our results show that switchable and enduring polarization in CGO films is governed by the electric field-driven redistribution of oxygen vacancies with a critical field strength of approximately 0.5 MV/cm at room temperature. This leads to notable and persistent pyroelectric effects with a coefficient of approximately 180 $\mu C\cdot m^{-2}\cdot K^{-1}$. Our findings highlight the potential to develop high-performance, sustainable, environmentally friendly polar film materials by manipulating ionic defects from their centrosymmetric ground states. This approach provides new opportunities to expand the range polar materials in current and future energy and electronic applications.**

*Keywords: defect-mediated polarization, induced pyroelectricity, oxide films, symmetry breaking*

## 1. Introduction

Symmetry breaking in crystalline materials – often related to structural or magnetic phase transitions - is fundamentally important and has garnered significant attention for producing new and unexpected phenomena across many scientific disciplines. A key example of such symmetry breaking is the occurrence of piezoelectricity in crystals when the inversion center is lost in phase transition.[1-5] Very often, more symmetry elements are additionally lost, leading to a polar order. These materials encompass not only piezoelectrics, but also pyroelectrics and ferroelectrics. Piezoelectricity refers to the ability of a material to interconvert mechanical and electric energies, exhibiting mechanically induced charge displacement (direct effect) and electrically induced mechanical displacement (converse effect).[1,6,7] Pyroelectricity is the temporal variation in spontaneous or built-in charge polarization under thermal cycling.[8-10] Ferroelectrics are a subset of pyroelectrics, usually characterized by a long-range co-operative interaction of entities generating spontaneous polarization that can be reversed by an external electric field.[1,6,7] Some materials that were historically thought to be only pyroelectric (or electrically polar) were recently found to have a switchable polarization after suitable ion substitution, and thus are ferroelectric.[11]

The distinction between piezoelectrics and pyroelectrics depends on the material symmetry, usually presented in terms of point groups of the crystal structure.[1] All classical piezoelectric materials require non-centrosymmetric crystal structure, while pyroelectricity appears only in a limited number of non-centrosymmetric materials with polar symmetry, which allows permanent net dipole moments in the absence of external mechanical or electric fields. Beyond pure monocrystal properties governed by point symmetries, the notions of a centricity and polarity can also be applied to polycrystalline, amorphous, and composite materials by extending point symmetries with Curie (or limiting) group symmetries.[12] This suggests the possibility of engineering piezoelectricity and pyroelectricity in materials that are not expected to exhibit these effects considering the point group symmetries of their crystal structure; a well-known example is electrets.[13,14]

Breaking the inherent inversion symmetry of materials is, therefore, a core concept to create emerging piezoelectric and pyroelectric properties that do not exist in the parent material. Such symmetry breaking has been achieved by applying external stimuli such as strain gradients and electric fields, as well as by modifying the material's composition.[3,4,15] It can also be achieved by engineering materials interfaces and heterostructures,[5,16] microscopic or macroscopic control of charged point defects and their ordering,[3,17] and chemical doping.[2,18] For instance, a recent work shows an effective defect-mediated symmetry breaking and the emergence of charge polarization by creating and regulating defect orderings in intrinsically centrosymmetric oxides through post-annealing treatments, resulting in non-centrosymmetric polar phases (e.g., $AgFeO_{2+\delta}$).[19] Of particular interest is the possibility of redistributing ionic defects within a material by controlling an electric field, e.g., moving charged defects towards the oppositely biased electrode. This process can effectively induce macroscopic asymmetry even though the sample's crystal structure may remain locally centrosymmetric. This approach imparts functional charge asymmetry to material through the control of defect dynamics, differing from symmetry breaking associated with crystal phase transitions that occur throughout the lattice of an entire material.

To explore such a field-driven defect redistribution and the ensuring symmetry engineering, cerium oxide ($CeO_2$) doped with Gd, $Gd_xCe_{1-x}O_{2-\delta}$ (CGO), and other aliovalent 3+ cations such as Pr, Sm, or Nd serves as a convenient model material due to its ability to host anion defects, oxygen vacancies ($V_O^{\bullet\bullet}$, which represents $V_O^{2+}$ in Kröger-Vink notation).[20] In such systems, two types of oxygen vacancies can be typically present: compensating $V_O^{\bullet\bullet}$, formed to maintain charge neutrality with the dopant cations, and excess $V_O^{\bullet\bullet}$, introduced during material processing, and charge-compensated by $Ce^{3+}$.[3] These material systems have shown the impact of dynamic $V_O^{\bullet\bullet}$ on the charge redistribution and mechanical deformation, thereby providing a platform to investigate field-induced symmetry breaking, its associated piezoelectric and pyroelectric effects. In addition, aliovalent $Gd^{3+}$ supplies localized magnetic moments while $V_O^{\bullet\bullet}$ stabilizes $Ce^{3+}/Ce^{4+}$ mixed valence, which could give rise to defect-mediated magnetic order.[21] Although long-range magnetic order in CGO and other doped $CeO_{2-\delta}$ has not been established, the combination of localized magnetic moments, mixed valence, local inversion-symmetry breaking, and effective spin–orbit coupling — often mediated by more delocalized 5*d*–O 2*p* hybridization or defect-induced 4*f*–5*d* mixing — could provide a basis for defect-mediated and emergent antisymmetric exchange interactions (e.g., Dzyaloshinskii–Moriya–type).[22] Such interactions could in turn open pathways to magnetoelectric coupling and electrically tunable magnetic behavior.

Recently, large electrostrictive responses of oxidized CGO and other doped ceria ceramics with $x \leq 15$ at.% were found in the order of $\sim 10^{-16}$ $m^2/V^2$ in the low-frequency regime (< 1 Hz), exhibiting dopant-independent behavior but strong measurement frequency dependence.[23] This indicates that the rate-dependent contributions can be attributed to the presence of mobile ionic species ($V_O^{\bullet\bullet}$) in the ceria samples. Furthermore, by introducing excess $V_O^{\bullet\bullet}$, large piezoelectric-like strains in the doped $CeO_{2-\delta}$ films have been demonstrated under controlling electric fields (e.g., simultaneous large static + relatively small alternating field applications) and also exhibit reversible response phase angles ($\Delta\theta = 180°$) with respect to the applied driving fields.[3]

Doped ceria originally possesses a centrosymmetric fluorite structure (space group: $Fm\bar{3}m$). It was found that the phase transition to a tetragonal structure ($P4_2/nmc$) can occur and become energetically stabilized when introducing a high $V_O^{\bullet\bullet}$ content (e.g., $\delta \approx 25\%$) into $CeO_2$ lattices. The dominant mobile charged defects are $V_O^{\bullet\bullet}$ and compensating electrons, which appear as 4$f$ electrons on Ce ions, resulting in a negatively charged $Ce_{Ce}'$ (small polaron).[24] The oxygen vacancy and polaron can form deformable and orientable electrical and elastic dipoles in the crystal structure. These two can be further oriented within unit cell level and moved by external fields over large distances (more than one unit cell) within the crystal. In general, dipoles, clustered defects, and space charges can lead to pyroelectric polarization and piezoelectric strain.[13]

Here, we demonstrate that it is possible to induce a persistent built-in charge polarization in the CGO films, primarily driven by the asymmetric distribution of $V_O^{\bullet\bullet}$ (and associated negatively charged defects, primarily $Ce_{Ce}'$), as illustrated schematically in **Figure 1a,b**. Our investigation indicates that the defect-associated macroscopic polarization in the CGO films is electrically tunable. By reversing the asymmetric distribution of $V_O^{\bullet\bullet}$ across the film layer, we can achieve a switching of both the induced polarization and the pyroelectric current. The pyroelectric performance of the electrically poled CGO films near room temperature (RT), with a coefficient of $\mu_P$ ~180 $\mu C \cdot m^{-2} \cdot K^{-1}$, is comparable to current commercial pyroelectrics such as $Pb(Zr,Ti)O_3$ (PZT) films, $LiNbO_3$ (LNO) and $LiTaO_3$ (LTO) single crystals. This work lays the groundwork for creating a new class of polar materials (which are both piezoelectric and pyroelectric) in ionic oxides by controlling and stabilizing the distribution of mobile ionic defects.

## 2. Results

### 2.1. Defect-mediated macroscopic built-in polarization and pyroelectricity in CGO films

In this work, the polarization of the CGO film samples was examined using a dynamic method to measure the pyroelectric current of the samples while modulating their temperature at a constant rate during the heating and cooling cycles (**Figure 1c**).[10] This method is convenient as it removes interference of thermally stimulated currents with the polarization signal.

Pyroelectricity originates from the variation of spontaneous polarization, $P$, with temperature, $T$, and is characterized by the pyroelectric compliance, $\mu_P = dP/dT$. The polarization change generates a compensating current in the electric circuit and in the sample, which is called pyroelectric current, $I_P$. The pyroelectric current thus requires a temporal variation of the sample's temperature, $dT/dt$, i.e., heating or cooling of the sample,[9,10] in contrast to a static, spatial temperature gradient required for the thermoelectric effect.[25] The relationship between the $I_P$, $\mu_P$, and $dT/dt$ is given by:

$$I_P = \mu_P \frac{dT}{dt} A, \quad (1)$$

where $A$ is the electrode area over which charges are collected. Therefore, assuming that the pyroelectric compliance is constant, the pyroelectric current as a function of time behaves as the time derivative of the temperature (also see **Figure S1** and **Section S1**, Supporting Information). To ensure homogeneous thermal stimulation of the samples, the frequency of the applied temperature triangular waveform was applied between 5 mHz and 50 mHz. The amplitude of temperature modulation ($\Delta T = T_{max} - T_{min}$) was varied around a given ambient temperature to determine the pyroelectric coefficient of the samples following Eq. (1).

The polycrystalline CGO (Gd 20 at.%) films were prepared with thicknesses ranging from approximately 0.4 to 1.8 µm using RF sputtering and pulsed laser deposition techniques. All films were deposited in an out-of-plane capacitance configuration on an electroded $SiO_2$(150 nm)/Si(500 µm) substrate at RT, followed by top electrode deposition. To ensure that the observed pyroelectric effect is not influenced by electrodes, different combinations of top and bottom electrodes were examined: Pt(~100 nm)/Cr(~5 nm)/CGO(~1.8 µm)/Al(~150 nm) and Al(~100 nm)/CGO(~1.8 µm)/Al(~150 nm) as shown in **Figure S2**, Supporting Information. Prior to the pyroelectric current measurement, the electrical resistance of both the as-deposited and electrically poled films was measured and determined to be ~50 – ~100 MΩ, corresponding to absolute current flow of 0.3 – 0.6 pA across the film samples in the pyroelectric measurement setup. Considering the large resistance of the CGO films, the leakage current is minute and thus the thermoelectric contribution is negligible. Details of the film preparation and pyroelectric measurements are further discussed in the Experimental Section, **Figure S1** and **Section S1**, Supporting Information.

Figure 1c shows the dynamic pyroelectric current density ($J_P = I_P/A$) of an as-deposited CGO film while modulating the temperature around ambient temperature at a cycling frequency of $f$ = 10 mHz. We observed a robust and stable dynamic pyroelectric current response in the CGO films measured at a constant temperature amplitude of $\Delta T$ = 3.7 °C and heating-and-cooling cycling rate around an ambient temperature of $T$ = 21 °C. A sign change in the current of the CGO film sample, which has the remanent macroscopic polarization oriented downward in the set-up (see **Figure 1b**), occurs during the transition from heating to cooling, that is, when $\frac{dT}{dt}$ changes the sign. A notable linear increase in the maximal reached current density $J_P$ with increasing $\Delta T$, in agreement with Eq. (1) ( $J_P \propto dT/dt$), is shown in **Figure 1d** and in **Figure S3**, Supporting Information. The calculated remanent polarization ($P_r$) change, defined as $P_r = \frac{Q}{A} = \frac{\int I_P dt}{A}$, where $Q$ denotes electric charge, follows the temperature increase and decrease (Fig. 1e). According to the sign of $J_P$ (positive for $\frac{dT}{dt} > 0$ and negative for $\frac{dT}{dt} < 0$), which has been centered around zero after subtracting the background current, it is concluded that the polarization within the CGO film is oriented toward the bottom electrode. We note that both the dielectric permittivity and pyroelectricity in the CGO film are considerably reduced (approximately one order of magnitude) after annealing in $O_2$ at 400 °C for 30 mins (**Figure S4**, Supporting Information). This indicates that the origin of the observed pyroelectricity is associated with the $V_O$ concentration in the film.

More direct evidence that the polarization in CGO is associated with a macroscopic asymmetric distribution of positively charged $V_O$ across the CGO film (from top to bottom, in an increasing concentration), as schematically illustrated in **Figure 1b**, will be presented in the next section. Note that the less-abrupt sign changes in the $J_P$ of the CGO film (compare the PZT signal in **Figure S5**, Supporting Information with that of CGO in **Figure 1c**) is most likely due to the slower kinetics of polarization change in the CGO film during the heating and cooling cycles. This was confirmed by measuring the pyroelectric current with different temperature cycling frequencies (**Figure S6**, Supporting Information). The slow polarization change also suggests an inhomogeneous defect distribution across the CGO film thickness and possibly the presence of sluggish polar entities within its crystal structure.

The occurrence of pyroelectric current in centrosymmetric as-deposited CGO film is unexpected. What are the possible mechanisms that can lead to it? The most general framework to address this question is the phenomenon of electrets.[13] In any sufficiently insulating dielectric that does not inherently possess a polar or non-centrosymmetric structure, polarization can emerge from the alignment of defect dipoles or an asymmetric distribution of charged defects within a dielectric, as illustrated in **Figure 1b** for the fluorite structure. In the case of electrets, the net non-zero macroscopic polarization is usually induced by applying a polarizing field, although spontaneous alignment of polar entities is known to occur in the special case of centrosymmetric precursor phases of ferroelectric materials.[26]

CGO contains charged mobile ionic defects, $V_O^{\bullet\bullet}$ and $Ce'_{Ce}$,[24] which can move and rearrange within the crystal. These defects can cluster to form "*dipoles*" (more precisely, dimers or trimers) that can be reoriented within the crystal by an electric poling field. Alternatively, individual defects (in this case $V_O^{\bullet\bullet}$ and $Ce'_{Ce}$) can be displaced over larger distances by an electric field, accumulating in the near-surface region of the material. It has been shown that, in CGO, regions with a high $V_O$ concentration can undergo a phase transformation from the cubic fluorite phase (*Fm-3m*) to a non-polar tetragonal phase (*P4₂/nmc*),[3,27] which further contributes to the heterogeneity of the structure that may be necessary for the generation of the pyroelectric response.[13] In the case of CGO films, we observe that defects tend to self-align or self-segregate in the absence of a poling field. While the reasons for this built-in polarization are not immediately apparent, it suggests that a poling field may induce even a stronger, stable, and long-lasting polarization state. In the following section, we will demonstrate that all these conditions are met in CGO films.

**2.2. Direct Observation of Asymmetric $V_O$ Distribution within the CGO film**

To clarify the correlation between the generation of the pyroelectric effect and the $V_O$ distribution in the CGO film, we conducted transmission electron microscopy (TEM) measurements. Details of these measurements are given in the Experimental Section, Supporting Information. **Figure 2a** displays electron diffraction (ED) patterns corresponding to the selected areas (1 – 7) of the CGO film layer, progressing from the top to the bottom region as depicted in the cross-sectional STEM image. Interestingly, our findings indicate a gradual change in the ED patterns when measured towards the bottom electrode (see the plots for the radial intensity integration and ratios of the first two diffraction peaks from zero reciprocal distance) as illustrated in **Figure 2b,c**. This reveals the phase change from the cubic fluorite phase (*Fm*-3*m*) to a new tetragonal phase ($P4_2/nmc$),[3,27] verified by our simulation of the ED patterns (see **Figure S7**, Supporting Information). Moreover, to understand how the phase transition correlates with the distribution of $V_O$, we conducted energy dispersive x-ray spectroscopy (EDX) measurements for elemental analysis of the CGO film (**Figure 2d**). Notably, a gradual increase in the $V_O$ ratio clearly appears towards the bottom areas of the film. (**Figure 2f**) In contrast, the distribution of Gd dopants (avg. ~22.5±1.2%) remained consistent across the entire film layer (**Figure 2e**). Note that the calculated $V_O$ ratio $[= 1 - (\frac{Z_O}{2(Z_{Ce}+Z_{Gd})})$, where $Z_A$ is the ratio (concentration in unit cell) of atom A] presented here is the absolute atomic ratio with respect to the stoichiometric cation and anion ratio of $CeO_2$ (1:2). For example, doping $CeO_2$ with 20 at.% $Gd^{3+}$ leads to formation of 5 at.% $V_O$ for charge neutrality. However, our analysis shows that a composition gradient of $V_O$ with local ratio up to ~20 at.% can be formed and stabilized within the CGO film (**Figure 2f**). This can be explained by the excess $V_O$ in CGO films, which are generated during material processing and are neutralized by electrons trapped at Ce, forming mobile small polarons, $Ce'_{Ce}$.[3,27,28] Our findings also reveal a direct correlation between the observed $V_O$ concentration gradient and the tendency for the phase transition within the film (**Figure 2c**). These results demonstrate that an asymmetric Vo distribution is indeed accompanied by polarization and pyroelectric effect in the as-grown films.

### 2.3. Polarization Switching in the CGO film by External Electric Field

Next, we investigated the tunability of asymmetric $V_O^{\bullet\bullet}$ distribution by applying static (poling) electric field ($E_{DC}$) to the CGO film. Firstly, dynamic pyroelectric measurements were carried out on a pristine CGO film, which possesses a built-in downward-oriented polarization, accompanied by a $V_O$ concentration gradient as shown in the previous section (a lower concentration of $V_O$ near the top surface and higher near the bottom electrode). A sufficiently large $E_{DC}$ (≈-0.8 MV/cm) was then applied on the top electrode, inducing polarization that is oriented opposite to that of the pristine film. The field was applied for different poling times, following the sequence of 15 mins → 45 mins → 60 mins → 120 mins. After each field application, the polarization behavior was continuously monitored by measuring the pyroelectric current. Each pyroelectric test was performed for more than 2 hours to ensure that each poling sequence resulted in a stable state of the film, as shown in **Figure S8**, Supporting Information.

**Figure 3a** shows variations in the dynamic $J_P$ of the CGO film as a function of the poling time with $E_{DC}$ = -0.83 MV/cm. The current was measured by applying a triangular heating-and-cooling temperature waveform at $f$ = 10 mHz under an average $T$ of 22 °C with $\Delta T$ = 2.4 °C. We observed a distinct sign change of $J_P$ for $t \geq 45$ mins. Additionally, there was a significant increase in the amplitude of $J_P$ by approximately six times with respect to the magnitude in the pristine film (see **Figure 3a,b**). These results reveal the switchability of the built-in polarization within the CGO film; the governing mechanism is most likely the redistribution of the positively charged $V_O$ and associated negative charges by the poling field. The field-induced $V_O$ redistribution across the film layer enhances polarization with respect to that in the pristine state, leading to a larger pyroelectric effect with an increasingly square-like shape of $J_P$.

The reversible polarization was further investigated by applying different strengths of the static poling field to the sample, while keeping a constant poling time of 120 mins. We found that a field of $E_{DC} \geq \approx\pm 0.5$ MV/cm is necessary for the polarization switching (**Figure 3c**). Lastly, we observed that the enhanced pyroelectric effect of the poled CGO films can be maintained for at least 6 hours without any significant degradation of the current amplitude (**Figure 3d,e**).

### 2.4. Persistent Symmetry Breaking, Polarization, and Pyroelectric effect

The observed endurance of the pyroelectric effect in the CGO films after electrical poling suggests a non-volatile macroscopic symmetry breaking of the electrically poled films. Thus, we also anticipate the emergence of the associated piezoelectric effect in the poled films. To verify this, we conducted converse piezoelectric measurements on a CGO film in a frequency range from 100 mHz to 1 kHz, measured under various external electric field conditions: (i) simultaneous application of a relatively large $E_{DC}$ of +0.5 MV/cm and a small driving $E_{AC}$ of 12 kV/cm (see **Figure S9**, Supporting Information), (ii) application of only $E_{AC}$, the piezoelectric strain measured 4 hours later after removing the poling $E_{DC}$, (iii) application of the same $E_{AC}$, with the strain measured 12 hours after removal of $E_{DC}$, and (iv) subsequent measurements with the same $E_{AC}$ one day and two days after removal of $E_{DC}$. Note that the positive $E_{DC}$ applied on the top electrode reinforces the built-in polarization. Moreover, it is worth mentioning that accurately determining the piezoelectric susceptibility of the pristine CGO films would be challenging while preserving the initial state by avoiding field-driven defect movements or rearrangement. It would require measuring the converse effects by applying a very low field ($E_{AC} \leq 12$ kV/cm), which would limit the detection of mechanical displacements (e.g., < 0.3 Å with $d_{33}$ of < 10 pm/V).

**Figure 4a** shows the frequency dependence of the apparent converse piezoelectric coefficient, $|d_{33}^*|$, of the CGO film, continuously measured under different field applications over 2 days. Firstly, we found a large piezoelectric effect ($|d_{33}^*| > 500 - 2500$ pm/V) of the film at a low frequency regime (< 100 Hz), excited by applying the combined $+E_{DC}$ and $E_{AC}$ field, as we reported previously.[3] The piezoelectric effect results from the $E_{DC}$-enforced built-in asymmetric distribution of charged defects throughout the film layer and electric-field biased electrostriction.[3,6] The large low-frequency $|d_{33}^*|$ (> 1000 pm/V) is attributed to rate-dependent contributions from asymmetrically distributed $V_O$, which should be the main polarization mechanism in these films. Note that $|d_{33}^*|$ is ~100 pm/V at 1 kHz, comparable to commercial PZT thin films. After removing the applied $E_{DC}$, we observed a decrease in the overall $|d_{33}^*|$ of the film, while measuring displacement with the same driving $E_{AC}$ (12 kV/cm) after a 4-hour interval without any field. The decrease is due to a partial loss of the induced polarization. Following the continuous electromechanical measurements, we observed the $|d_{33}^*|$ of the CGO remained nearly constant after 12 hours and up to 2 days: $|d_{33}^*|$ was approximately 150 pm/V at 100 mHz and approximately 50 pm/V at 1 kHz, still respectable values. These results directly confirm the persistence of macroscopic symmetry breaking and charge polarization of CGO film, which is a consequence of the $E_{DC}$-reinforced asymmetric distribution of $V_O$ across the film.

To further verify polarization robustness, we conducted dynamical pyroelectric tests on the same poled samples (**Figure 3d,3e**) after one year. Remarkably, we observed the same large pyroelectric response from the aged samples with nearly consistent performance. **Figure 4b** illustrates the dynamic $J_{\mathrm{P}}$ of a CGO film, poled by a $+E_{\mathrm{DC}}$ (= 0.8 MV/cm), and after ageing for one year upon the field removal. The pyroelectric response was also measured by applying two different heating-and-cooling waveforms (triangular- followed by a square-shaped) at $f$ = 20 mHz, at an average $T$ of 21.5 °C with $\Delta T$ = 2.4 °C in the triangular temperature waveform ($\Delta T$ = 2.6 °C in the square waveform) (also see **Figure S10**, Supporting Information). The two different waveforms were used to verify that the pyroelectric response indeed follows Eq. (1), where current should behave as the time derivative of the temperature change. For the triangular temperature waveform, the aged CGO film exhibits a nearly square-shaped $J_{\mathrm{P}}$, which was well-defined with relatively swift sign-switching behavior (e.g., $\frac{dT}{dt} > 0 \rightarrow +J_{\mathrm{P}}$ to $\frac{dT}{dt} < 0 \rightarrow -J_{\mathrm{P}}$. A steep increase-and-decrease in $J_P$ was observed in response to the square temperature waveform as one would expect from Eq. (1) for the derivative of the square function where $T$ rapidly raises in time ($\frac{dT}{dt} \gg 0$ or $\frac{dT}{dt} \ll 0$). Furthermore, we observed that the amplitude of $J_P$ linearly varies with $\Delta T$ from 0.3 to 2.8 °C, following the $J_{\mathrm{P}} \propto dT/dt$ relationship described by Eq. (1) (**Figure 4c,4d**). The corresponding pyroelectric coefficient, $\mu_{\mathrm{P}}$, is found to be approximately 180 $\mu C \cdot m^{-2} \cdot K^{-1}$. In addition, the CGO samples exhibit highly stable pyroelectric performance at elevated average temperatures (up to approximately 70 ºC) without notable degradation. (see **Figure S11**, Supporting Information). All these results confirm that the charge polarization of the poled CGO behaves similarly to that in common pyroelectrics and, after an initial relaxation upon the field removal, remains robustly stable for at least one year.

## 3. Discussion

Our results demonstrate that as-deposited CGO films, which possess a centrosymmetric fluorite structure, unexpectedly exhibit a durable and spontaneous macroscopic symmetry breaking revealed by a built-in polarization. We present direct evidence that the charge polarization is associated with an asymmetric distribution of oxygen vacancies. The formation of asymmetric $V_O$ concentration within CGO films occurs during the film preparation. Moreover, the polarization can be tuned and switched by electrically controlling the distribution of mobile $V_O$. As shown in **Figure 3**, the field-induced $V_O$ redistribution enhances the overall polarization relative to the pristine state, resulting in a larger pyroelectric effect. The key factor behind the observed pyroelectricity is the macroscopic distribution of $V_O$ and the microscopic orientation of dipolar defects, i.e., $Gd^{3+}$–$V_O$ and $Ce^{3+}$–$V_O$ pairs, during the poling process.[3] The microscopic dipoles, which are partially aligned by electric poling, respond to temperature changes through reversible interatomic distance variations, contributing to the pyroelectric current. Meanwhile, a significant $V_O$ concentration (≥ 20 %) is asymmetrically redistributed across the CGO film, forming stable space charge regions. Near the negative electrode, a large space charge region is formed due to the accumulation of positively charged $V_O$ and notably, this persists even after the electric field is removed. $Gd^{3+}$ ions (Gd′), which remain immobile, create a negative space charge region in areas where $V_O$ are depleted. This charge separation leads to the formation of a macroscopic dipole. When excess $V_O$ is present in CGO, charge compensation is driven by $Ce^{3+}$ ions (Ce′) which contain additional 4*f* electrons (small polarons). The activation energy for small polaron transport via electron hopping is comparable to that for $V_O$ diffusion.[24] The small polarons could accumulate near the positive electrode during the poling process. This negative space charge further contributes to the macroscopic dipole. The observed pyroelectric current arises from changes in the film thickness with temperature. As the thickness increases (or decreases) with rising (or decreasing) temperature, the separation between the positive and negative space charge regions changes, thereby contributing to the generation of pyroelectric current. Thus, this defect-mediated polarization becomes further enhanced and stabilized after electric poling.

The emergent piezoelectric and pyroelectric effects in the CGO films are stable even after one year of ageing. A component of the observed macroscopic polarization could be potentially linked to a macroscopic strain gradient that accompanies the vacancy concentration gradient, presumably giving rise to a static flexoelectric polarization.[4] The flexoelectric polarization is difficult to quantify in such a complex system (e.g., nonlinear $V_O^{\bullet\bullet}$ gradient) (see **Figure 2f** and **Figure S9** in Supporting Information) and partial phase transition across the film layer (**Figure 2a** and **Figure S7** in Supporting Information ) and this contribution should be further explored. Of crucial importance for the control of oxygen, i.e., $V_O$ migration in CGO at or near RT by electric field, is the excess $V_O$, which are not bound to $Gd_{Ce}$.[29] The reason is that the $V_O$ bond with aliovalent dopants such as Gd has a much higher activation energy for migration (~0.75 - 1 eV) than those vacancies that are compensated by $Ce_{Ce}^{3+}$ (< 0.5 eV) and which are formed during deposition under oxygen-poor conditions.[28] This excess $V_O$ concentration can be controlled by film preparation conditions such as deposition in oxygen-reduced atmosphere.[28] Optimal $V_O$ concentration needs to be determined for effective defect and property engineering. The most intriguing observation is that a large $V_O$ concentration (20 at.% or more) on one side of the film can be effectively stabilized in the $CeO_2$ lattice, leading to polarization without considerable defect relaxation over a long time. This stability contrasts with significant thermodynamic charge relaxation associated with $V_O$ distribution in $SrTiO_{3-\delta}$ and $TiO_{2-\delta}$ observed after the removal of external stimuli, e.g., mechanical force and electric fields, leading to the gradual disappearance of the induced piezoelectricity and incipient ferroelectricity.[30,31] The $V_O$-rich parts of the CGO films, however, do undergo a phase transition, which may be required for the stabilization of space charges (macroscopic dipole).

The discovered ability of CGO to maintain asymmetry in the distribution of charged defects over a long period presents great opportunities for developing new and unexpected functional properties. Thus, CGO films exhibit durable and robust pyroelectric performance with $\mu_P$ ~180 $\mu C \cdot m^{-2} \cdot K^{-1}$, which is comparable to or even higher than the coefficients of present commercial lead-based and lead-free pyroelectrics, e.g., PZT films (~130 – ~200 $\mu C \cdot m^{-2} \cdot K^{-1}$), LTO and LNO (~80 and ~180 $\mu C \cdot m^{-2} \cdot K^{-1}$) single crystals.[8,32] Furthermore, the persistent macroscopic symmetry breaking and pyroelectric properties observed in CGO films outperform those in other defect-engineered polar materials (e.g., $V_O$-induced polar phase (*P*4*mm*) in a $SrTiO_3$ single crystal, which exhibits $\mu_p$ ~30 $\mu C \cdot m^{-2} \cdot K^{-1}$ at RT under simultaneous polarizing field applications).[33] From a technological perspective, pyroelectric polar thin films are of particular interest for fabricating one- and two-dimensional arrays for thermal sensing, imaging, and gas detection in the frame of micro-electro-mechanical system (MEMS) technology.[8] The use of polar CGO film materials with engineered charge polarity offers the significant advantage of simpler, lower-temperature processes (near RT), making them fully compatible with Si-based complementary metal-oxide semiconductor (CMOS) technology. This contrasts sharply with the longstanding disadvantage of directly integrating ferroelectric thin films (e.g., Ti-rich PZT) onto read-out chips, which typically require high processing temperatures, e.g., 700 °C.[8] Hence, our discovery provide new opportunities for developing a new class of highly sustainable, biocompatible polar materials – achieved through defect engineering in fluorite oxides – and fosters new avenues for technological advancements in practical and future piezoelectric and pyroelectric film applications.

## 4. Experimental Section

### Material preparation and characterizations

In this work, CGO films with thicknesses of approximately 0.380, 1.25, and 1.80 μm were prepared at RT using two different deposition techniques: RF magnetron sputtering and pulsed laser deposition (PLD). For sputtering, the film deposition was carried out at an RF power of 200 W with an Ar gas flow of 15 sccm and a working pressure of $1 – 2 \times 10^{-2}$ mbar. For PLD, a KrF excimer laser (λ = 248 nm) was employed with a laser fluence of ~2 $J/cm^2$, a repletion of 3 Hz, and a working pressure of ~1 × $10^{-6}$ mbar. In both depositions, a CGO (20 % Gd) ceramic target was used. Prior to film deposition, bottom Al electrodes (100 – 150 nm) were sputtered onto $SiO_2$ (200 nm)/Si(100) substrates at RT. Following CGO film deposition, top electrodes were deposited, consisting of either Pt/Cr (Pt: ~100 nm; Cr: 10 nm) or Al (~150 nm). The polycrystalline CGO films were characterized for microstructure, crystallinity, and composition using X-ray diffraction (Bruker D8 Advance; λ = 1.54056 Å), atomic force microscopy (Cypher VRS), TEM (Double Cs-corrected ThermoFisher Scientific Titan Themis 60–300), and EDX. EDX spectroscopy was performed on the electron microscope operated at 300 kV with a beam current of 100 pA, acquisition was done for approximately 20 minutes using a SuperX detector and the Bruker Esprit software. A sputtered CGO film (approximately 1.2 µm thick) was measured (**Figure 2**). The selected area electron diffraction (SAED) patterns were acquired with a 50 µm selected area aperture in parallel beam illumination and using a camera length of 115 mm. EDX analysis revealed Cr impurities below 5 % in sputtered CGO films, with a measurement uncertainty of 1 – 2 %. Cr-free CGO films (~380 nm thick) were deposited by PLD on Al (~100 nm)/$SiO_2$/Si substrates at room temperature. Comparison of sputtered and PLD films showed negligible effect of Cr contamination on their electromechanical and pyroelectric behavior.[3]

**Electromechanical and pyroelectric characterizations**

To determine electric field-mechanical displacement responses of the CGO films, a photonic sensing system (MTI-2100 Fotonic Sensor) was employed. An alignment stage was used to mount the samples, and mechanical displacements were measured from the reflective sample surface (the grounded top electrode) in noncontact mode while applying electric voltages to the samples. An electric voltage was applied to the bottom electrode, which was connected to either a voltage source meter or a lock-in amplifier, along with a voltage amplifier. The photonic sensor, with an effective probe area of approximately 0.79 $mm^2$, produces a voltage output that varies linearly with the distance between the sample and sensor and is used to convert this voltage into mechanical displacement, with a sensitivity of 7.565 × $10^{-6}$ meters per volt. During electromechanical displacement measurements, all the real-time output amplitude signals were concurrently recorded by using both a lock-in amplifier and an oscilloscope (Tektronix, MDO3014). Prior to measuring the CGO film samples, corrections for external offset voltages, electrical noise, and instrument sensitivity were carefully performed and validated using reference samples, including standard quartz crystals and PZT (Pz29) specimens, within the system. Dynamic pyroelectric current measurements were performed with a home-made setup. The temperature of samples is modulated using a Peltier element in a regular periodic fashion while concurrently monitoring the associated current waveforms. The temperature of samples was varied in different waveforms (e.g., triangular, square, and sinusoidal) by $\Delta T$, centered around a base temperature at a given frequency (3 - 50 mHz). The pyroelectric current was determined by measuring the output voltage through a resistor with sub-picoampere resolution. The details are given in Figure S1 and Section S1, Supporting Information.

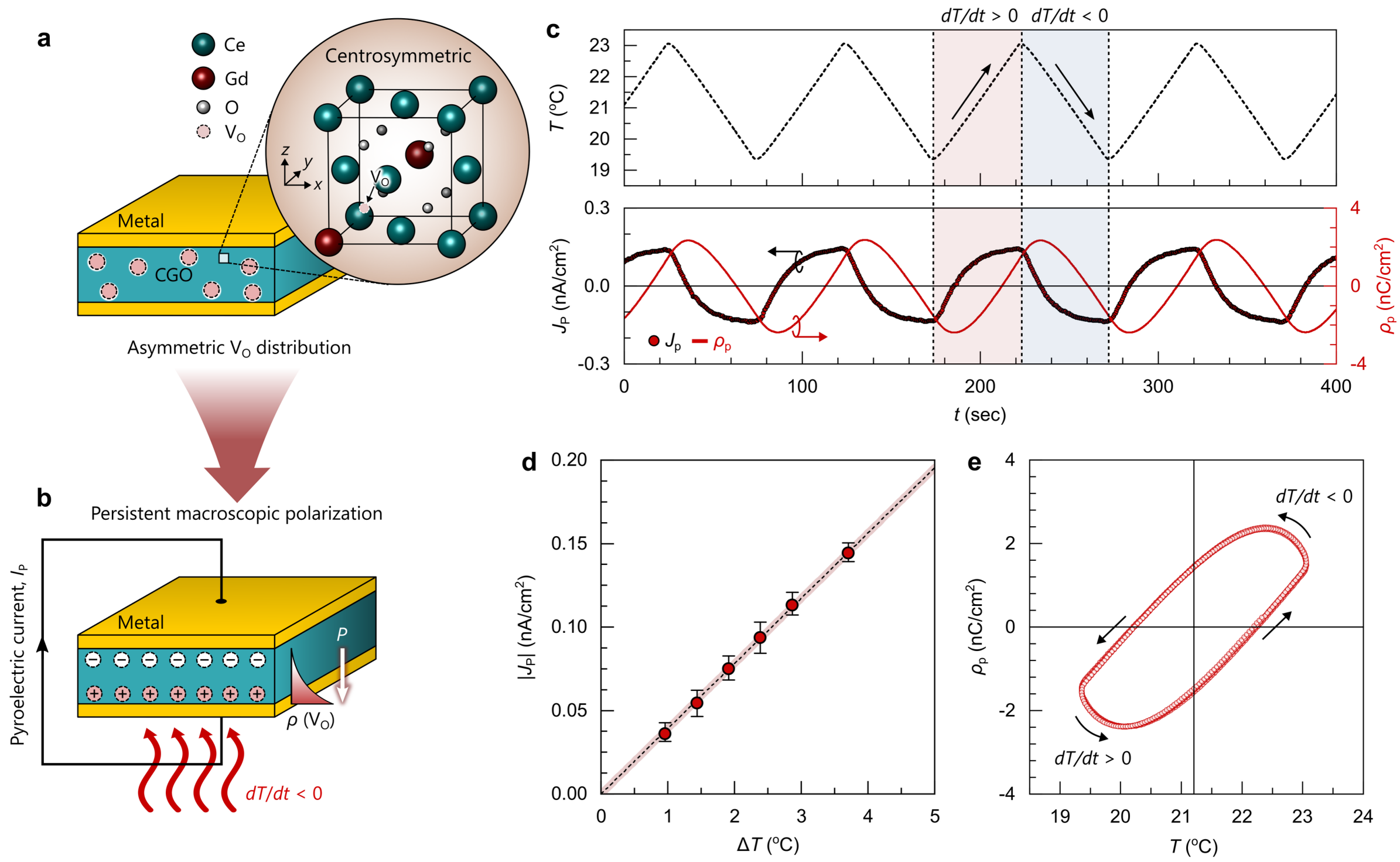

**Figure 1. Generation of pyroelectricity in the centrosymmetric CGO film by an asymmetric distribution of charged defects. a,b)** Schematic illustrations for **(a)** centrosymmetric CGO film layer with randomly distributed $V_O$ and **(b)** the formation of macroscopic built-in polarization in the CGO film. The polarization can be established through a macroscopic asymmetric distribution of charged point defects, $V_O$ across the film layer and the alignment of $V_O^{\bullet\bullet}$-$2Gd'_{Ce}/2Ce'_{Ce}$ clusters. **c)** Dynamic pyroelectric response of a CGO film sample, measured by continuously applying a triangular temperature waveform at a fixed cycling frequency of $f$ = 10 mHz under an average temperature of 22 °C with Δ$T$ = 3.7 °C. The upper panel shows the triangular temperature waveform applied to the sample at a fixed cycling frequency of $f$ = 10 mHz, with the heating ($dT/dt > 0$) and cooling ($dT/dt < 0$) parts of the cycle indicated. The lower panel shows the corresponding pyroelectric current density ($J_P$, left) and charge density ($\rho_P$, right) of the CGO. Note that CGO exhibits a large background current and $J_P$ is the variable component of the total current numerically centered around the zero. **d)** A linear variation in the maximum $J_P$ of the sample as a function of Δ$T$ following the relationship $J_P \propto \Delta T$. **e)** The temperature dependence of $\rho_P$ of the CGO film sample during the dynamic heating and cooling cycles. The hysteresis indicates a slow kinetics of polarization variation of the CGO sample.

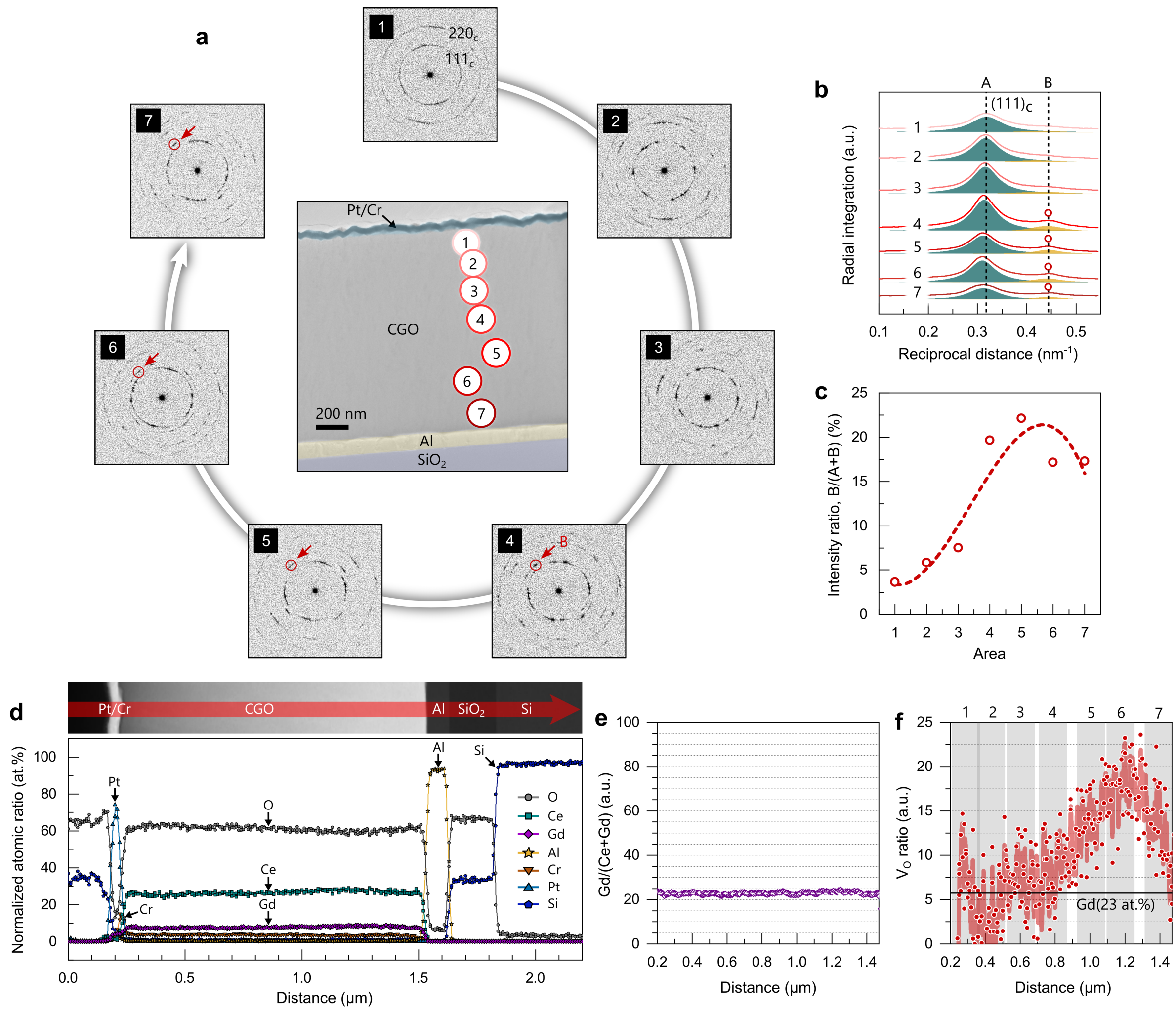

**Figure 2. Asymmetric distribution of oxygen vacancies in an as-grown CGO film layer. a)** A cross-sectional bright field TEM images of the as-deposited CGO film sample, Pt(100 nm)/Cr(5 nm)/CGO(1.2 µm)/Al(150 nm)/$SiO_2$(200 nm)/Si, presented in the center. The ED patterns (1 – 7) correspond to the local areas (1 – 7) of the CGO film. **b)** The integrated radial intensity profiles for the ED patterns (1 – 7). A Shift in the peak (A) and the appearance of new peak (B) are found from the top film surface (1) to the bottom area (7). **c)** A plot for the intensity ratio of the peak (B) to the total (A+B), [B/(A+B)]. **d)** EDX line profiles of the CGO sample along the arrow, denoted in the top cross-sectional high-angle annular dark field (HAADF) scanning transmission electron microscope (STEM) image. **e)** The atomic ratio of Gd to Ce in the CGO film, determined to be 22.8±1.6 %. **f)** A ratio profile of $V_O$ distribution across the film, determined by $[1-(Z_O/2(Z_{Ce}+Z_{Gd}))]$. The top electrode is on the left.

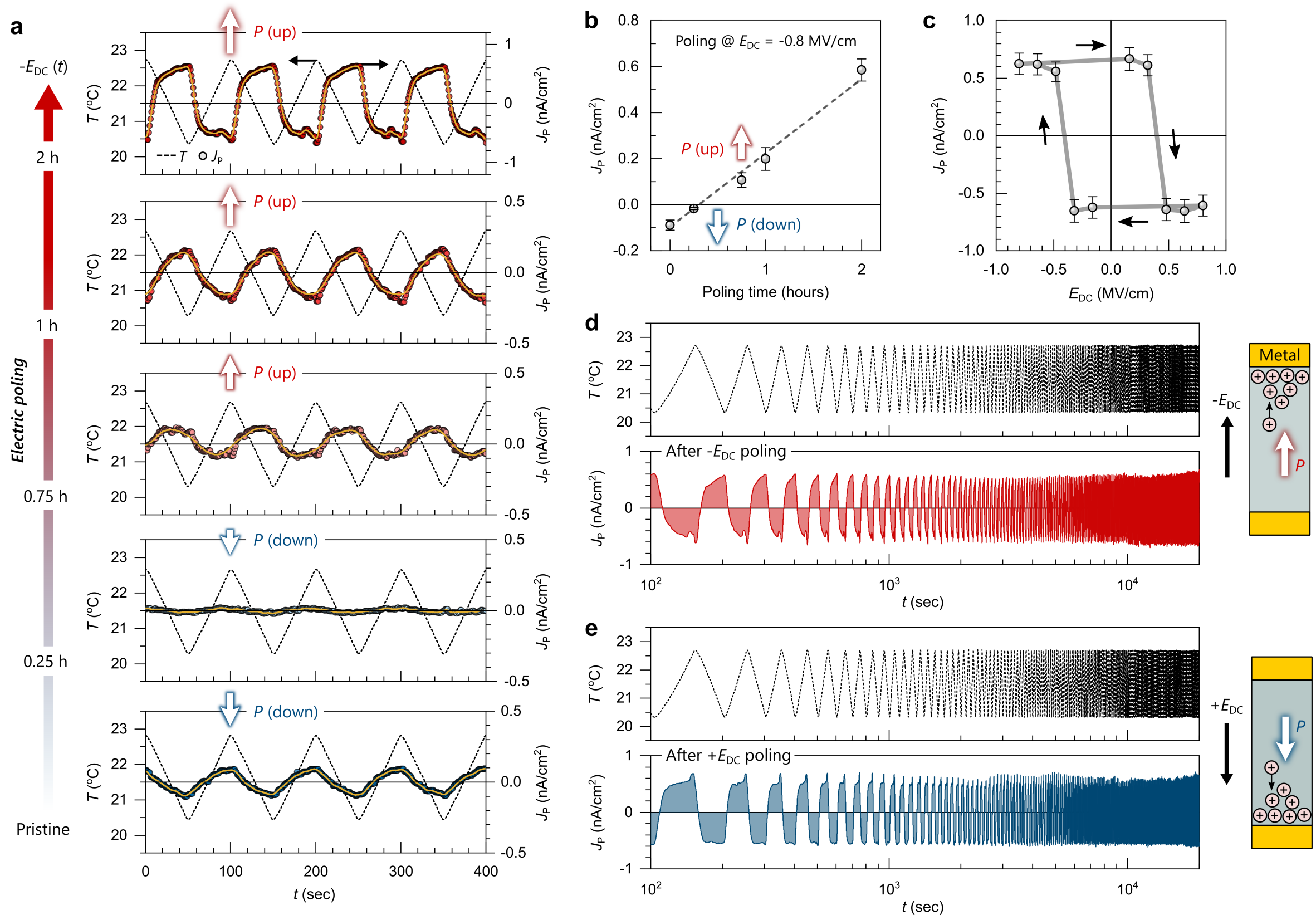

**Figure 3. Switching of pyroelectric current in a CGO film by electric poling. a)** Dynamic $J_P$ of the CGO film with a sequence of electric poling time (pristine(unpoled) → 15 min → 45 min → 60 min → 120 min), measured during triangular heating and cooling cycles at f = 10 mHz, under an average *T* of 22 °C with Δ*T* = 2.4 °C. For each poling sequence, a constant static field ($E_{DC}$) of -0.83 MV/cm was applied to the sample. **b)** Variation in the maximal $J_P$ of the sample as a function of poling time, measured during the cooling cycle with a constant rate of $dT/dt \approx$ -0.05 °C/sec (cycling at 10 mHz). **c)** Variation in the maximal $J_P$ of the sample, measured during the cooling process with a constant rate of $dT/dt \approx$ -0.05 °C/sec, illustrating the current switching. The pyroelectric response was measured after different $E_{DC}$ applications, followed by a complete sequence (-0.83 MV/cm → +0.17 MV/cm → +0.33 MV/cm → +0.5 MV/cm → +0.83 MV/cm→ -0.17 MV/cm → -0.33 MV/cm → -0.5 MV/cm → -0.83 MV/cm). Each $E_{DC}$ was applied for 2 hours. **d,e)** Performance endurance for the dynamic $J_P$ of the samples with continuous heating and cooling cycles, recorded for more than 6 hours after applying opposite poling fields, $E_{DC}$ = -0.83 MV/cm (**d**) and +0.83 MV/cm (**e**). In the right side, the schematic represents that the negative (positive) $E_{DC}$ moves positive charges towards the top (bottom) area of the CGO film, resulting in a macroscopic up (down)-polarization.

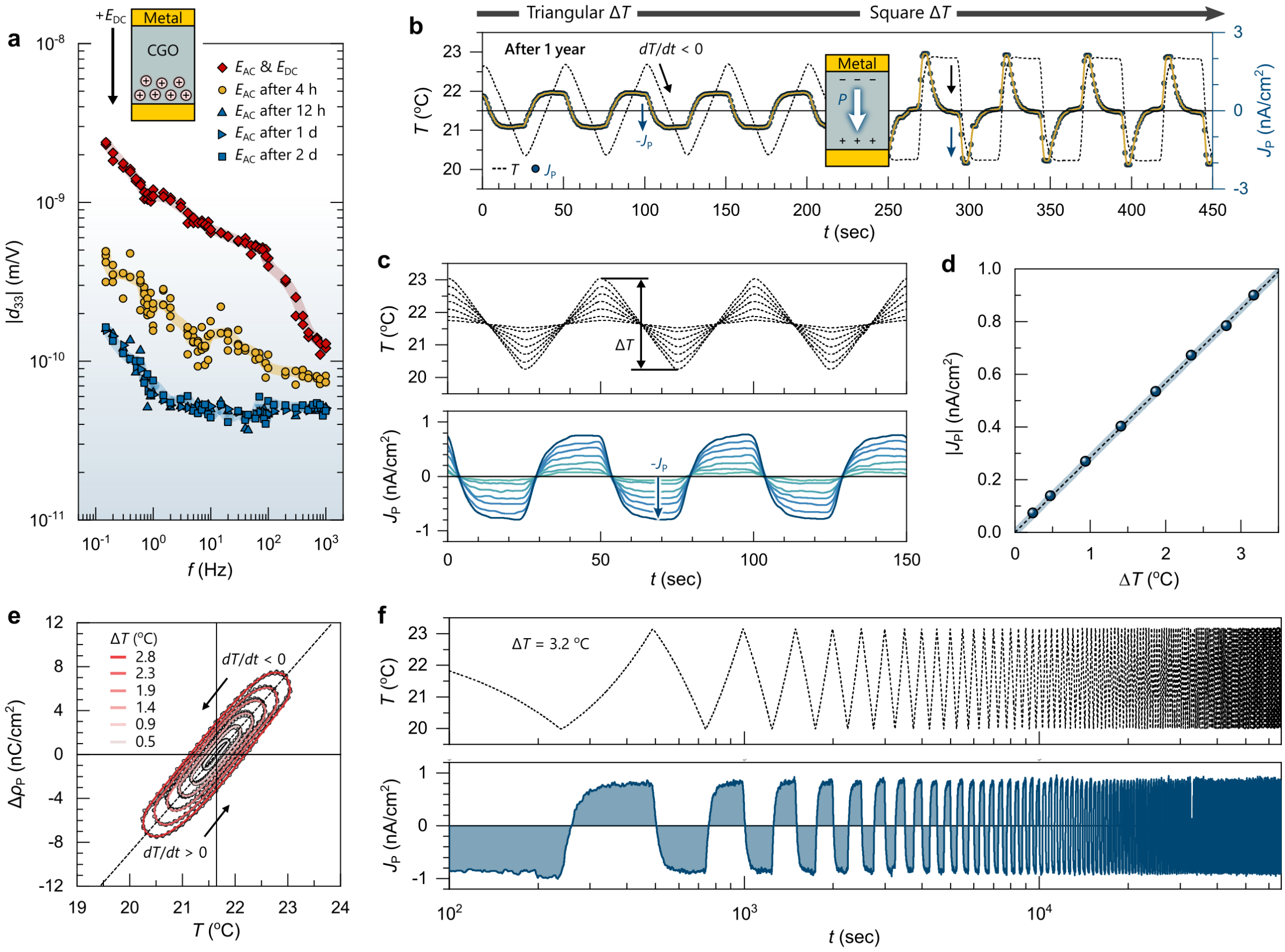

**Figure 4. Persistent piezoelectric and pyroelectric effects of the electrically poled CGO films. a)** The first harmonic electromechanical susceptibility, $|d_{33}|$, of a CGO sample as a function of $f$ (100 mHz ≤ $f$ ≤ 1 kHz) under different field conditions: (*i*) by concurrently applying a relatively large static DC field ($E_{DC}$ = +0.5 MV/cm) and a small driving AC field ($E_{AC}$ = 12 kV/cm), denoted by red rhombi, (*ii*) by applying only the driving $E_{AC}$, 4 hours after $E_{DC}$ was removed, denoted by yellow circles, and (*iii,iv,v*) by applying the same $E_{AC}$ 12 hours (bluish triangle) (*iii*), 1 day (bluish left-angled triangle) (*iv*), and 2 days (bluish square) (*v*) after $E_{DC}$ was removed. **b)** Dynamic $J_P$ of the poled CGO sample measured by continuously applying different temperature waveforms (triangular → square) of heating-and-cooling cycles, measured at $f$ = 20 mHz with an average $T$ of 21.5 °C and $\Delta T$ = 2.4 °C (2.6 °C) for the triangular (square) temperature waveform. The measured CGO sample was aged for approximately one year after poling. **c)** Dynamic $J_P$ of the aged CGO sample during the triangular heating and cooling cycles with different $\Delta T$. **d)** A linear variation in the maximal $J_P$ of the sample as a function of $\Delta T$. **e)** The variation of dynamic $\rho_P$ of the aged sample during the heating and cooling cycles. **f)** A prolonged $J_P$ measurement of the aged CGO sample under continuous application of heating and cooling cycles with $\Delta T$ = 2.4 °C for more than 18 hours.

# Supporting Information

## Tunable and Persistent Macroscopic Polarization in Nominally Centrosymmetric Defective Oxides

Dae-Sung Park,[1,2,3]* Nini Pryds,[1] Nicolas Gauquelin,[4] Mahmoud Hadad,[3] Dmitry Chezganov,[4] Alessandro Palliotto,[1] Daen Jannis,[4] Jorge Íñiguez-González,[5,6] Johan Verbeeck,[4] Paul Muralt,[3] Dragan Damjanovic[2]*

1. *Department of Energy Conversion and Storage, Technical University of Denmark, Fysikvej, 2800 Kongens Lyngby, Denmark.*
2. *Group for Ferroelectrics and Functional Oxides, Swiss Federal Institute of Technology–EPFL, 1015 Lausanne, Switzerland.*
3. *Group for Electroceramic Thin Films, Swiss Federal Institute of Technology–EPFL, 1015 Lausanne, Switzerland.*
4. *Electron Microscopy for Materials Science (EMAT), University of Antwerp, B-2020 Antwerpen, Belgium.*
5. *Luxembourg Institute of Science and Technology, 5 Avenue des Hauts-Fourneaux, L-4362 Esch/Alzette, Luxembourg.*
6. *Department of Physics and Materials Science, University of Luxembourg, 41 Rue du Brill, L-4422 Belvaux, Luxembourg.*

**E-mail: daepa@dtu.dk ; dragan.damjanovic@epfl.ch*

**Section S1: Instrumental setup and pyroelectric current measurements**

For pyroelectric tests, the pyroelectric current of samples was measured by implementing an in-house instrument that combines an electrometer, an analog PID temperature controller, and a power supply. In this work, all the current measurements were done in the absence of voltage bias. Samples were located on a copper sample stage with concentric edges to prevent secondary pyroelectric contributions (e.g., constant stress) caused by sample clamping. The temperature of the sample stage was periodically controlled with a thermoelectric element (Peltier elements) and was directly read by a T-type thermocouple, which sends accurate feedback voltages to the analog PID temperature controller. To avoid external effects (e.g., light exposure and temperature gradient), a polytetrafluoroethyene (PTFE) cover was used to enclose the samples during all the measurements. A Pt100 RTD was attached to the edge of the sample stage and used in a four-point configuration with a multimeter (Hewlett Packard, HP3478A) to measure the sample temperature.

To measure the current, we utilized a home-made electrometer compromising an operational transimpedance amplifier and a feedback resistor ($R_F$). The operational amplifier was connected to a sample to obtain an output voltage ($V_{out}$). The $V_{out}$ was measured by a multimeter (Hewlett Packard, HP3478A) and recorded by a LabVIEW program to read the current ($i_P = V_{out}/R_F$). This system utilizes the dynamical temperature oscillation method: the temperature of samples is modulated in a regular periodic fashion while concurrently monitoring the associated current waveform. The temperature of samples was straddled in various waveforms (e.g., triangular, square, and sinusoidal) by $\Delta T$ centered around a base temperature at a given frequency (1 - 50 mHz). The detailed experimental setup was described by Daglish.[10] In the short-circuit setup, the surface charge was measured as pyroelectric current from the surface of samples. Detailed information on the sampling, pyroelectric measurement, and electronic setups is schematically given in **Figure S1**. To verify the pyroelectric measurement system used in this work, we measured the pyroelectric current response of a prototypical poled ferroelectric, a commercial PZT ceramic (see **Figure S5**).

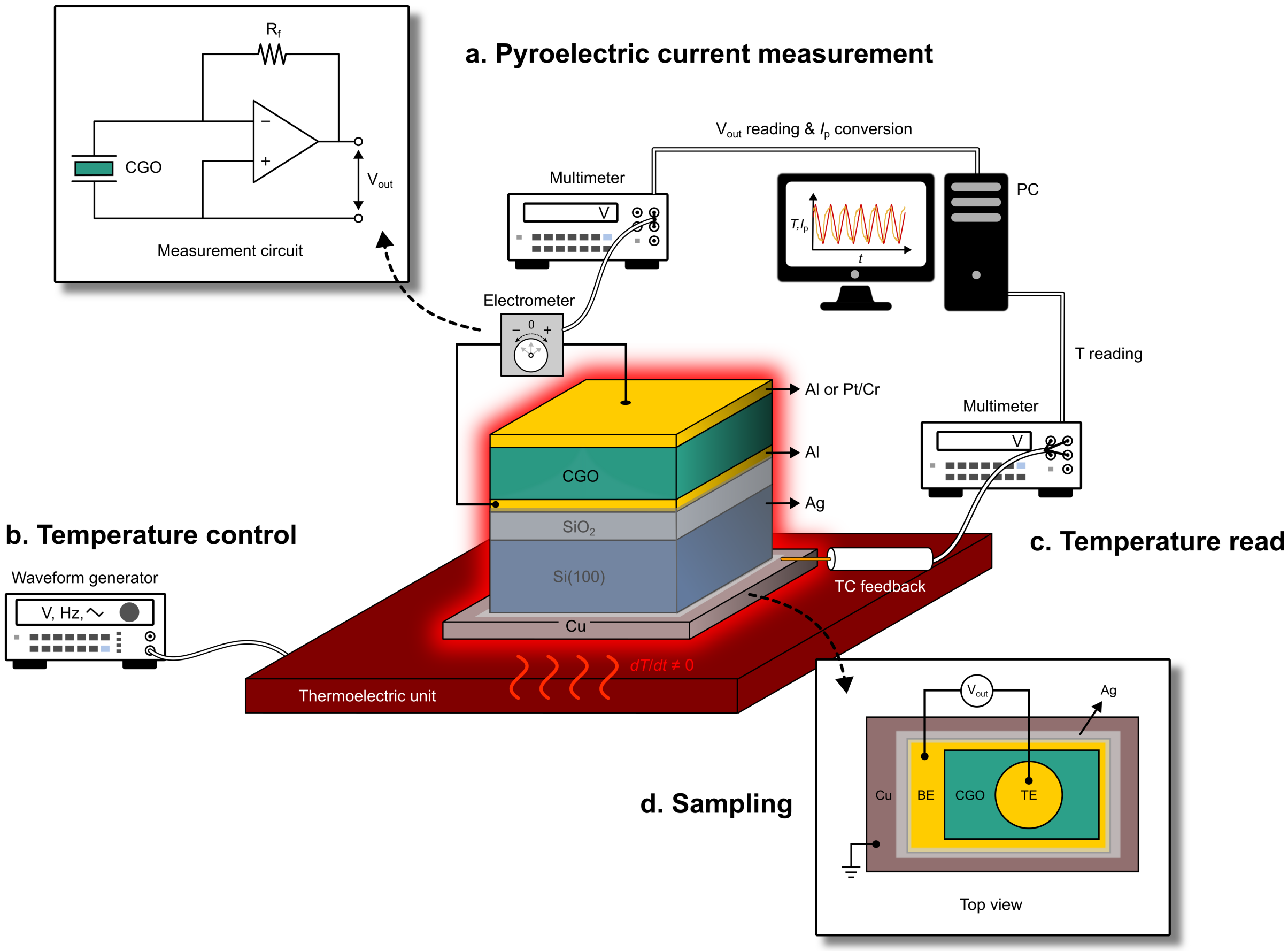

**Figure S1. Instrumental setups and sampling for dynamic pyroelectric measurements in time.** a) For pyroelectric current measurement, we utilized a home-made electrometer compromising an operational transimpedance amplifier and a feedback resistor ($R_f$). The operational amplifier was connected to a sample to obtain an output voltage ($V_{out}$). The $V_{out}$ was measured by a multimeter (Hewlett Packard, HP3478A) and recorded by a Labview program in a laboratory PC to read the current ($i_P = V_{out}/R_f$) as schematically shown in the measurement circuit. b,c) For temperature control (b) and read (c), the temperature of the sample stage was periodically controlled with a thermoelectric element (Peltier elements). The thermoelectric element was connected to an arbitrary waveform generator (Hewlett Packard, HP33120A) to set the temperature modulation parameters and waveforms (triangular, square, and sinusoidal). The temperature of the sample stage (Cu) was directly read by a T-type thermocouple (TC), which provided accurate feedback voltages to the analog PID temperature controller. The feedback voltages were measured in a four-point configuration using a multimeter (Hewlett Packard, HP3478A) to assess the sample temperature. The temperatures were concurrently recorded in the Labview program during the pyroelectric current measurements. d) For sampling, CGO samples consist of top (Pt/Cr or Al) and bottom (Al) electrodes, deposited on $SiO_2$/Si substrates. The top electrode of the samples was electrically connected with a probe while the bottom electrode was grounded. To avoid non-unform temperature control (heating and cooling), all the side areas of the $SiO_2$/Si substrates were covered by a metal, Ag, which was directly connected to the metal sample plate, Cu. All the samples were located on a copper sample stage with concentric edges to prevent secondary pyroelectric contributions (e.g., constant stress) caused by sample clamping. To avoid external effects (e.g., light exposure and temperature gradient), a polytetrafluoroethyene (PTFE) cover was used to enclose the samples during all the measurements.

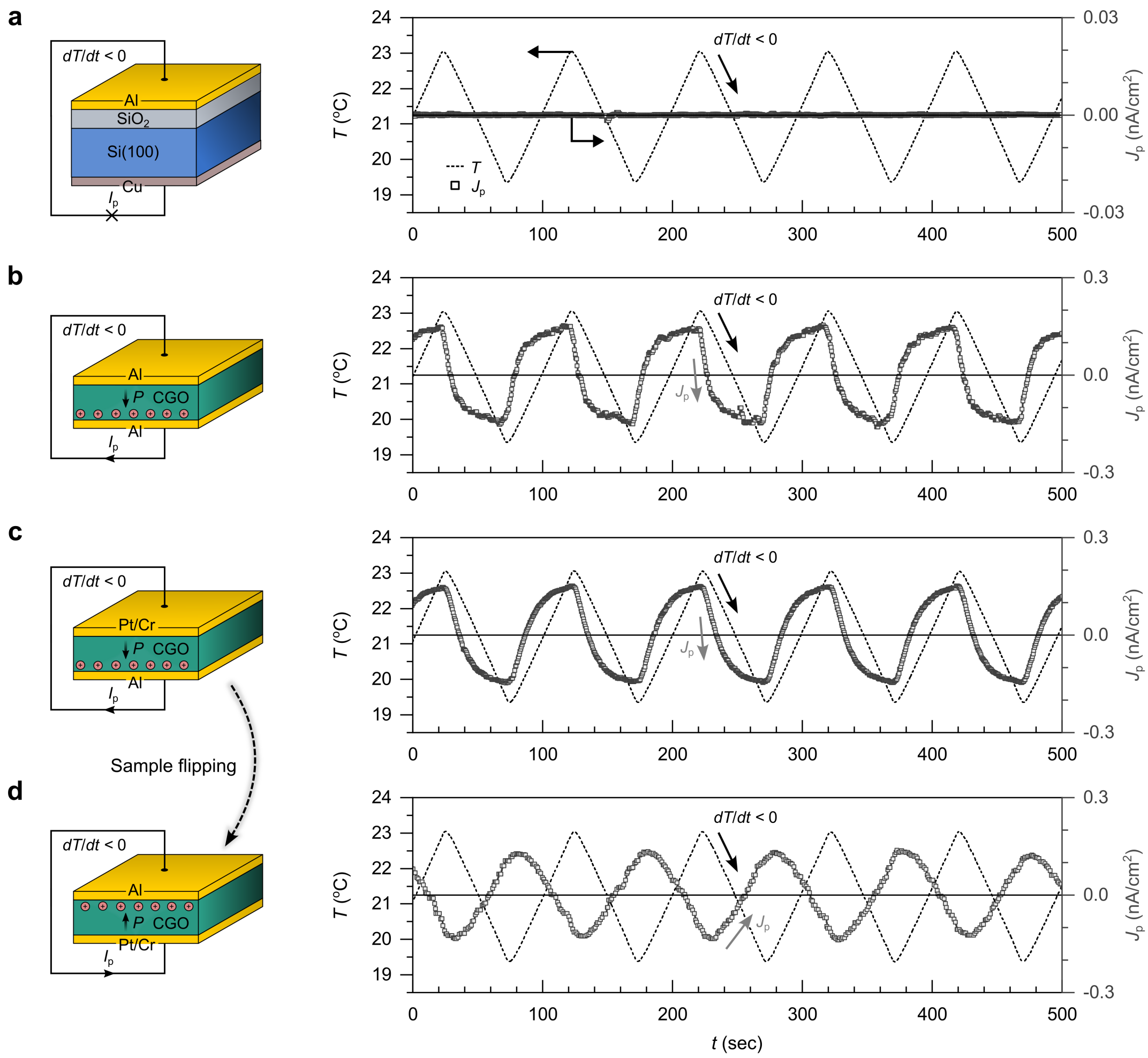

**Figure S2. Effects of metal electrodes sampling on the pyroelectric current density ($J_P$) of the CGO film samples.** a) Dynamic $J_P$ of a $SiO_2$/Si(001) sample which has the top (Al) and bottom (Cu) electrodes. No pyroelectric current occurs from the sample. b) Dynamic $J_P$ of a CGO film sample with symmetric top (Al) and bottom (Al) electrodes. c) Dynamic $J_P$ measurement of a CGO film sample with asymmetric top (Pt/Cr) and bottom (Al) electrodes. d) Dynamic $J_P$ of a CGO film sample with asymmetric electrodes after sample flipping. All the samples were measured under consistent temperature modulations by applying a triangular temperature waveform at $f$ = 10 mHz under an average set $T$ of 21.2 °C with $\Delta T$ = 3.7 °C. From this comparison, we observed no notable electrode effect (e.g., Schottky barrier effect) for the generation of pyroelectric effect in the CGO samples. In addition, the sign of $J_P$ reversed when the sample was flipped (see the schematics in (c) and (d). Note that the slower $J_P$ response of the flipped sample might be due to a non-uniform temperature modulation during the current measurement as the top surface of the sample was not perfectly connected to the Cu sample stage. Note that the current was centered around zero by removing the background current.

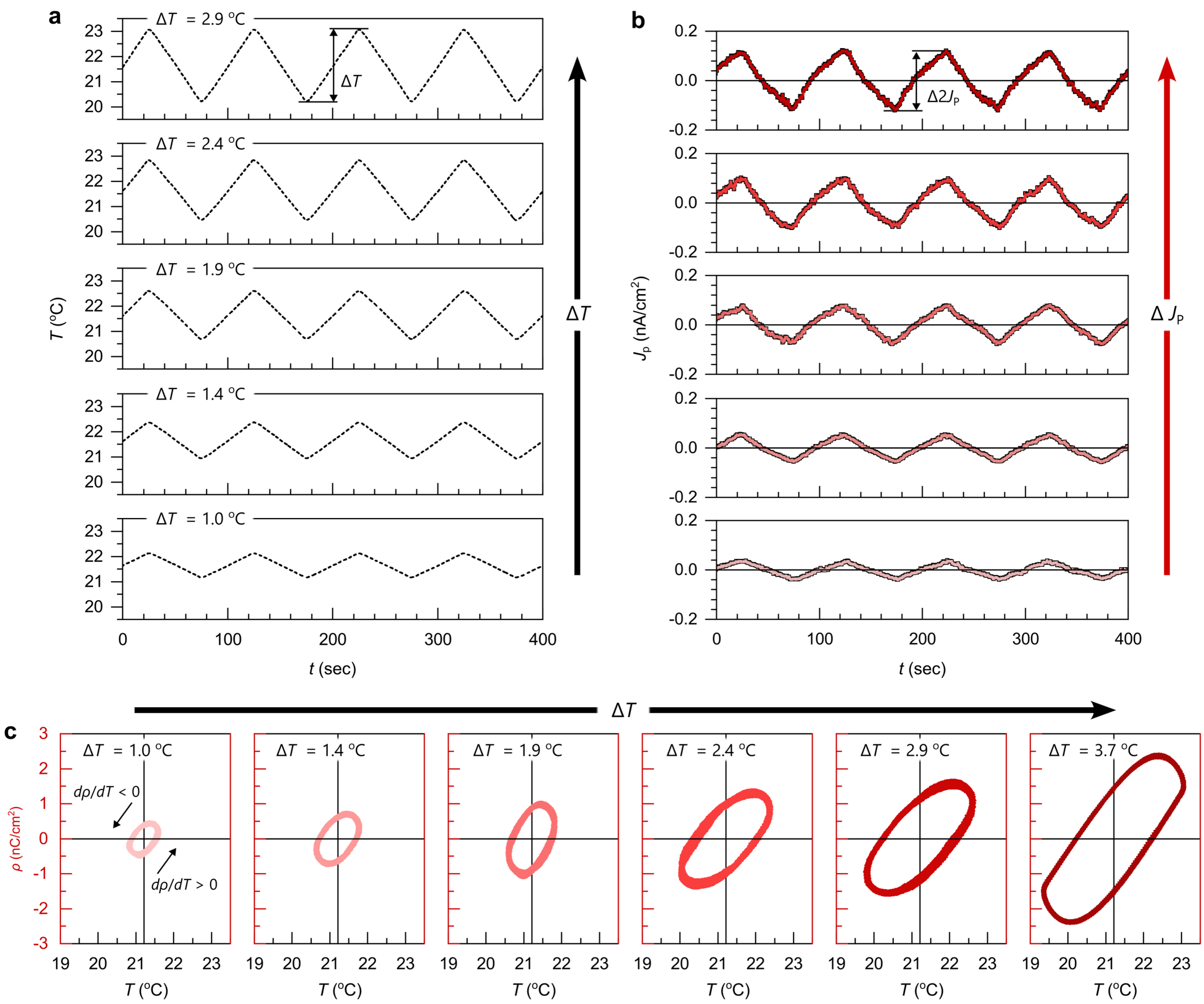

**Figure S3. The temperature amplitude ($\Delta T$) dependence of pyroelectric $J_P$ and charge density ($\rho_P$) of the CGO sample.** a,b) the dynamic $J_P$ of the CGO film sample as a function of $\Delta T$ (0.96 – 2.87 °C), measured with triangular heating and cooling waveforms across a set base $T$ of 21.6 °C. c) The corresponding $\rho_P$ of the sample. The CGO sample exhibits slow kinetic characteristics for the charge polarization switching once the heating ($dT/dt > 0$) or cooling ($dT/dt < 0$) starts. This reflects dispersive loops of $\rho_P$ during the temperature modulations. The current in (b) was centered around zero by removing the background current.

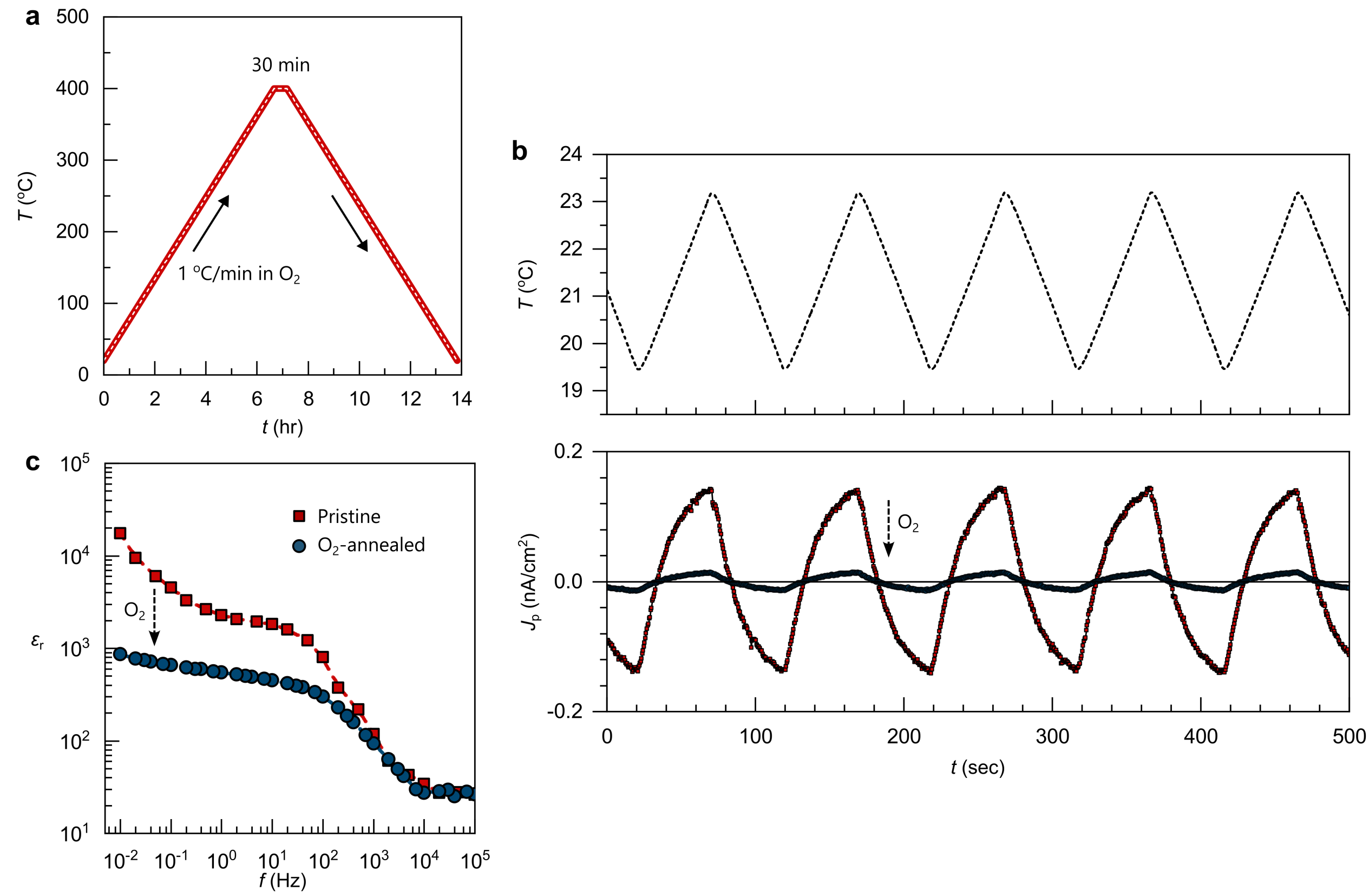

**Figure S4. $O_2$-annealing effect on the dielectric permittivity and pyroelectric $J_P$ of the CGO film samples.** a) The sequence of thermal annealing of a CGO film sample in an $O_2$ flow. The $O_2$ annealing was carried out at 400 °C for 30 mins with a temperature ramping rate of 1 °C/min. b) Comparison for the pyroelectric $J_P$ of the CGO sample before and after the $O_2$ annealing. The upper panel shows the applied triangular temperature waveform in time (a set base $T$ of 21.2 °C, $\Delta T$ = 3.7 °C, and $f$ = 10 mHz). The lower panel shows the corresponding $J_P$ of the sample before and after the $O_2$ annealing. c) The dielectric permittivity ($\varepsilon_r$) of the CGO film before and after the annealing. Both the $\varepsilon_r$ and pyroelectric effects of the CGO sample significantly decreased after the $O_2$ annealing according to the pyroelectric coefficient, $\mu_P = dP/dT = d\varepsilon E_{in}/dT$. This directly indicates that $V_O$ is the key component to generate the pyroelectric effects in the CGO system. The current in (b) was centered around zero by removing the background current.

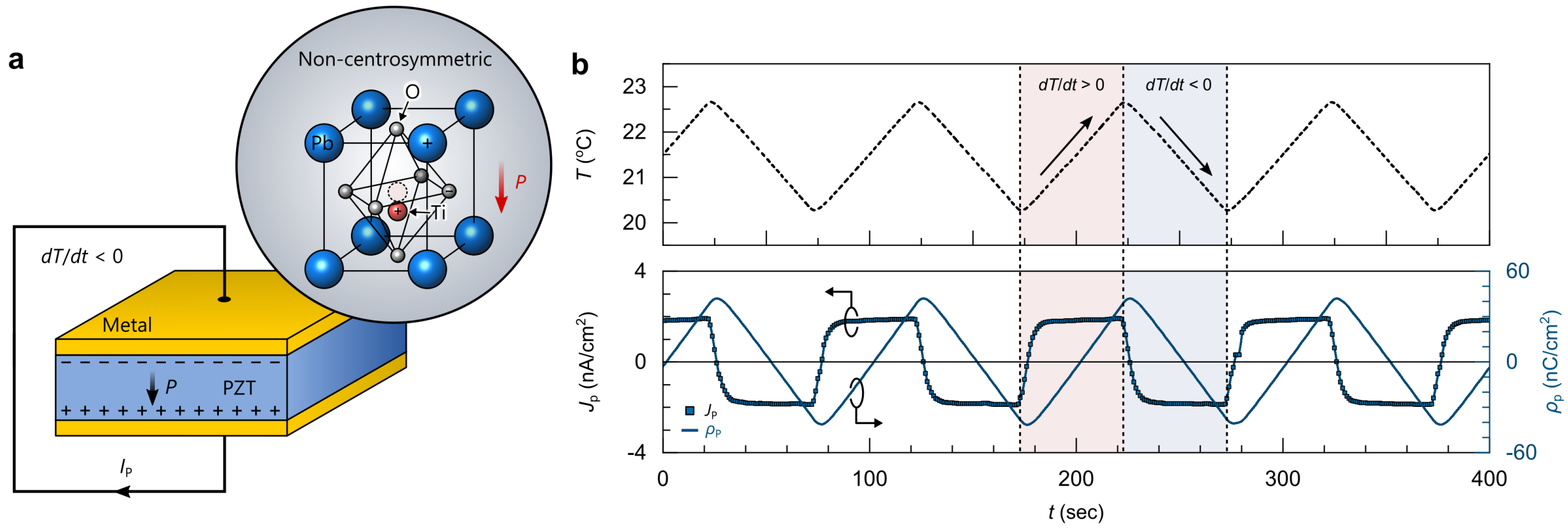

**Figure S5. Pyroelectric response of a noncentrosymmetric PZT ceramic.** a) Schematic of spontaneous net polarization in a conventional non-centrosymmetric, polar material, the perovskite ferroelectric PZT. Note the displacement of the Ti ion toward one O ion inside the octahedral cage. This off-centering of Ti together with shifts in position of other ions contributes to electrical polarization of PZT. The polarization of PZT, which has a regular $ABO_3$ perovskite structure (a), is built from electronic and ionic contributions arising from hybridization of electronic states and ionic interactions and displacement of Pb (A-site), Zr/ Ti (B-site) and O ions (see as an example Ti off-centering in a tetragonal $PbTiO_3$ unit cell along the [001] direction in (a).[S1] In conventional pyroelectrics/ferroelectrics, such local and long-range cooperative processes within the framework of the ordered structure give rise to the formation of polarization which can be modulated by temperature.[S2] b) Dynamic pyroelectric response for a reference PZT ceramic. An abrupt sign change in the current of the PZT sample, which has the remanent macroscopic polarization oriented downward in the set-up (a), occurs during the transition from heating to cooling, that is, when $\frac{dT}{dt}$ changes the sign. The current in (b) was centered around zero by removing the background current.

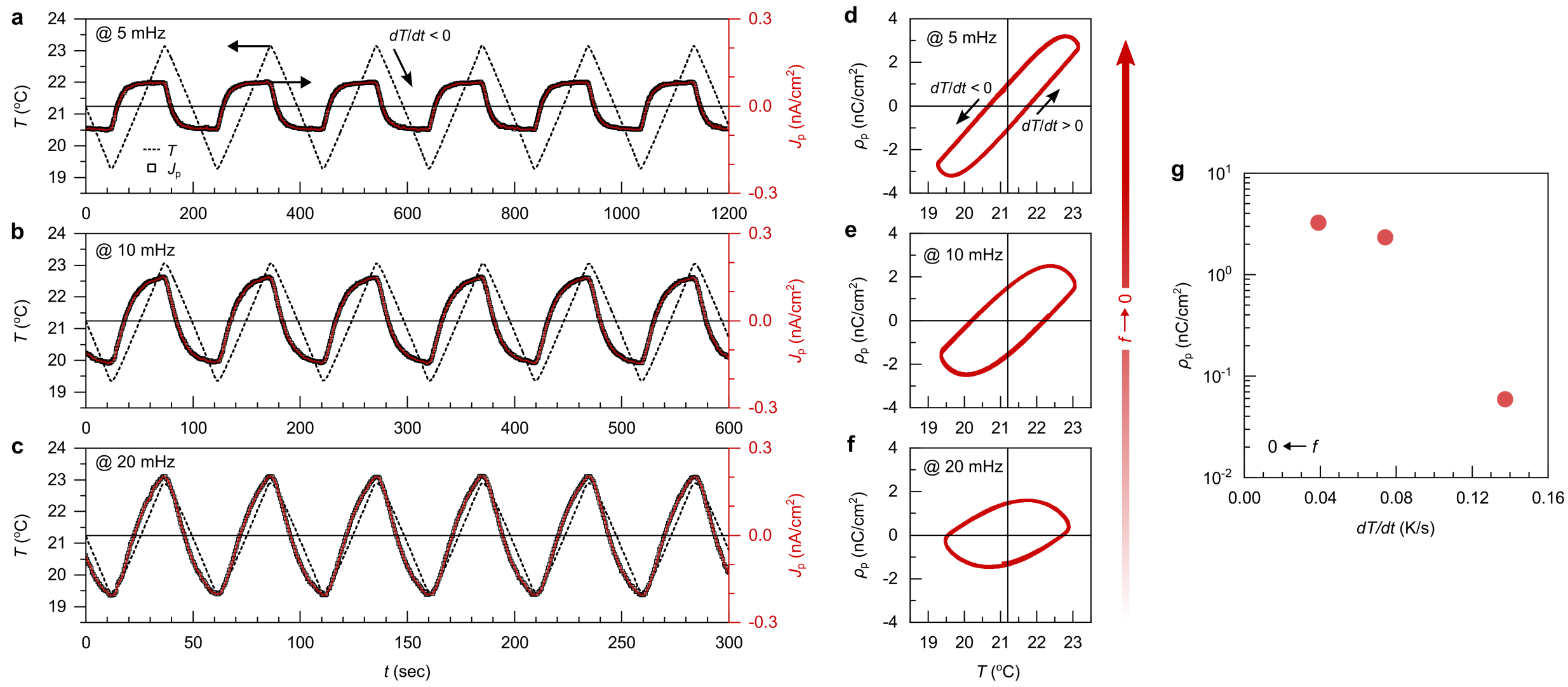

**Figure S6. Rate-dependent pyroelectric $J_P$ and $\rho_P$ of the CGO sample.** a,b,c) the dynamic $J_P$ of the CGO film sample as a function of the frequency, $f$ = 5 mHz (a), 10 mHz (b), and 20 mHz(c), of the triangular heating-and-cooling cycles, measured at a constant set base $T$ of 21.2 °C and $\Delta T$ = 3.7 °C. d,e,f) The corresponding $\rho_P$ of the sample during the heating-and-cooling cycles at $f$ = 5 mHz (d), 10 mHz (e), and 20 mHz(f). g) A plot for the $\rho_P$ versus $dT/dt$ of the sample. These results confirm the slow kinetic of charge polarization switching in the CGO film sample. The current in a-c was centered around zero by removing the background current.

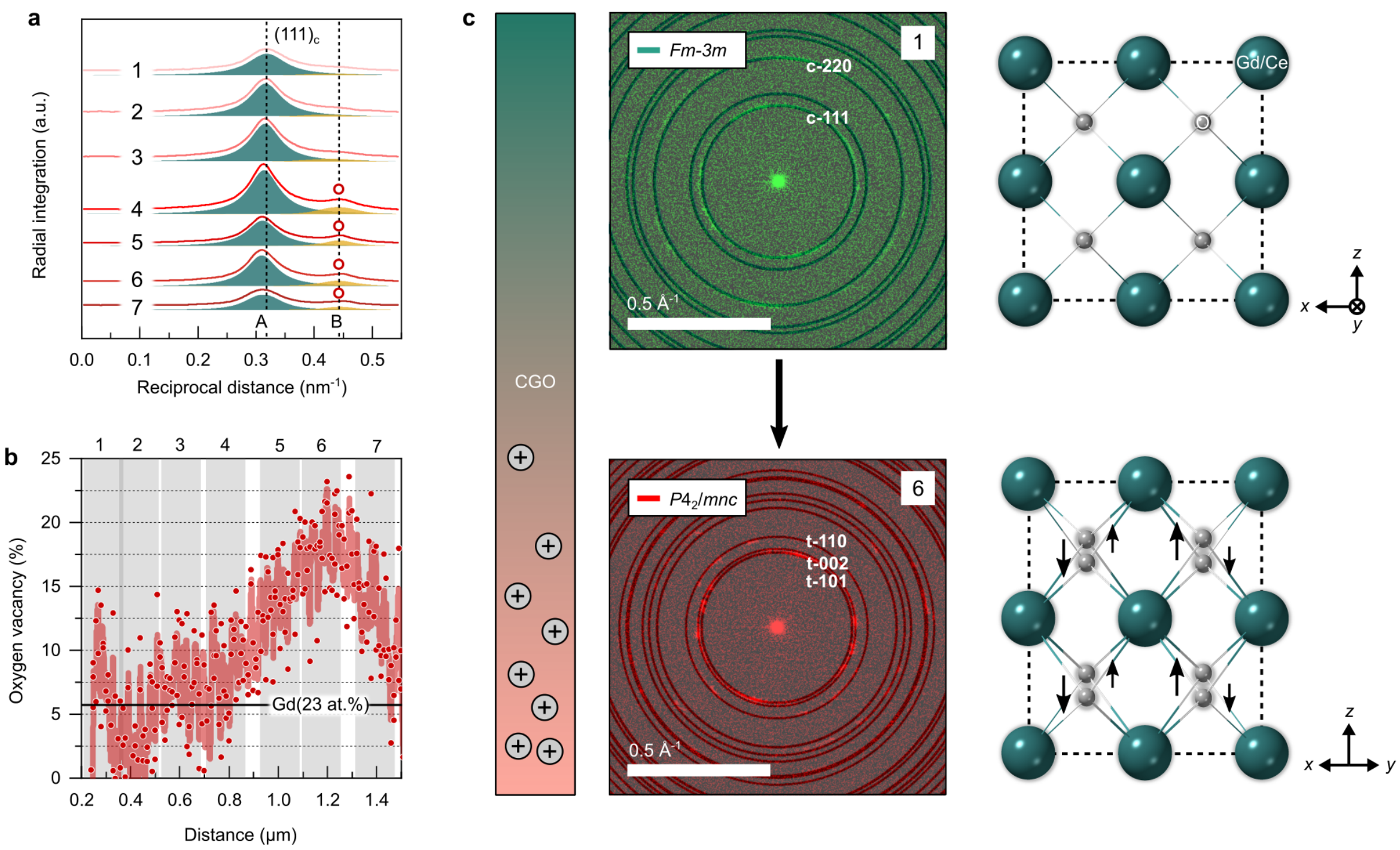

**Figure S7. $V_O$-induced phase transformation of the CGO film.** a) The integrated radial intensity profiles for the ED patterns (1 – 7) of the CGO film, presented in Fig. 2 of the main text. The numbers 1 to 7 correspond to the local areas within the CGO film layer from the topmost surface (just below the top Pt/Cr electrode) to the bottommost area (adjacent to the bottom Al electrode). b) A ratio profile of $V_O$ distribution across the film layer. c) Comparison between the experimental and simulated (line rings) ED patterns for the two different regions (1 and 6) of the film layer. In the region (1), a typical $V_O$ concentration was found to be approximately ~5 – 6 %, with respect to the Gd 20 - 22.5 %. The corresponding experimental ED pattern matches a simulated pattern with a cubic fluorite CGO (*Fm*-3*m*). In contrast, the region (6) exhibits a significantly higher $V_O$ concentration, approximately ~20 %. The corresponding ED pattern in this region matches a simulated pattern with tetragonal phase (*P*4$_2$/*mnc*).

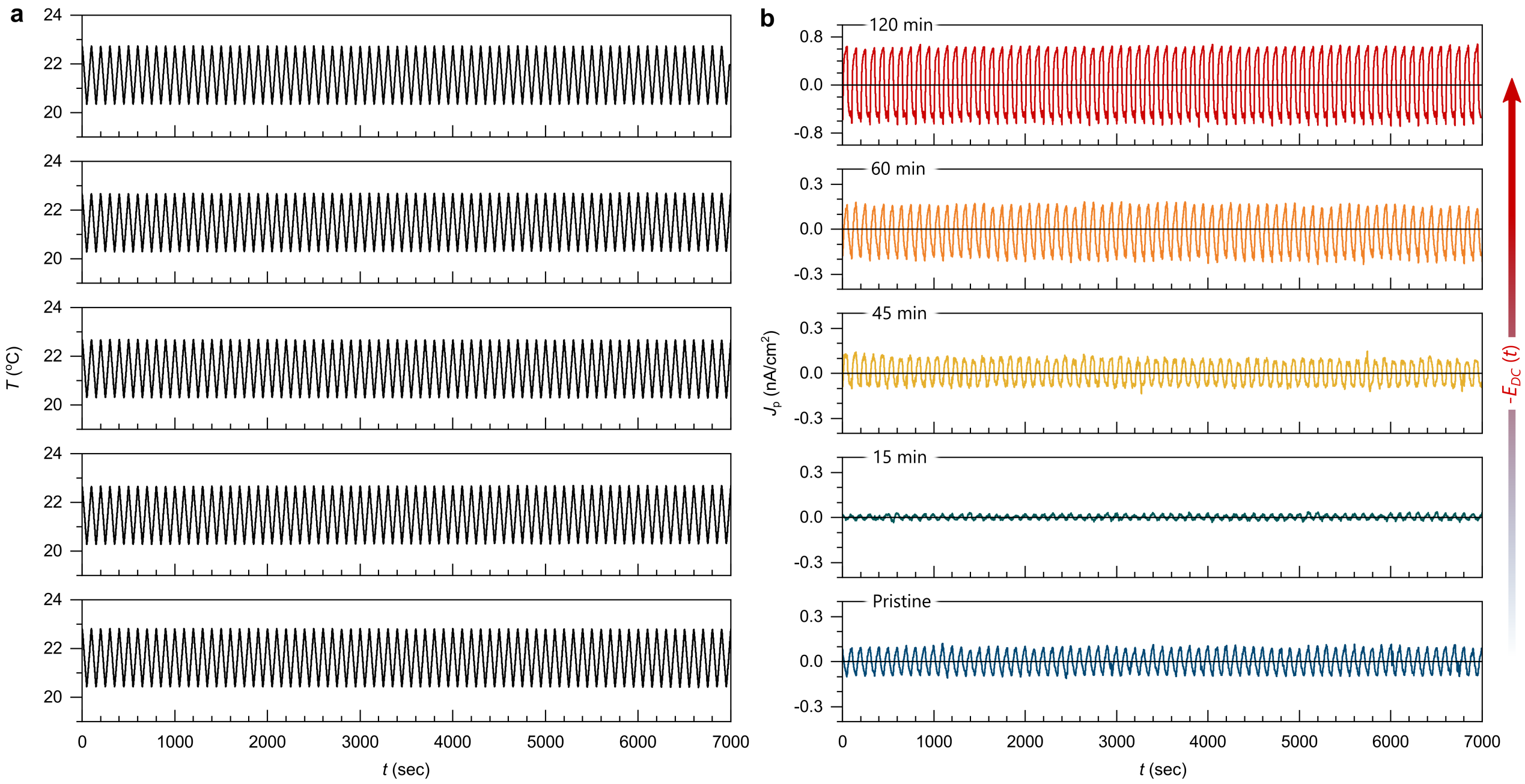

**Figure S8. Variations in pyroelectric $J_P$ of a CGO film sample as a function of electric poling time.** a) Constant triangular heating-and-cooling waveforms over time, applied to the CGO sample, with a base set $T = 21.5$ °C, $\Delta T = 2.4$, and $f = 10$ mHz. b) The corresponding $J_P$ of the sample, measured after each static field ($E_{DC}$) application, following a sequence of pristine → 15 min → 45 min → 60 min → 120 min.

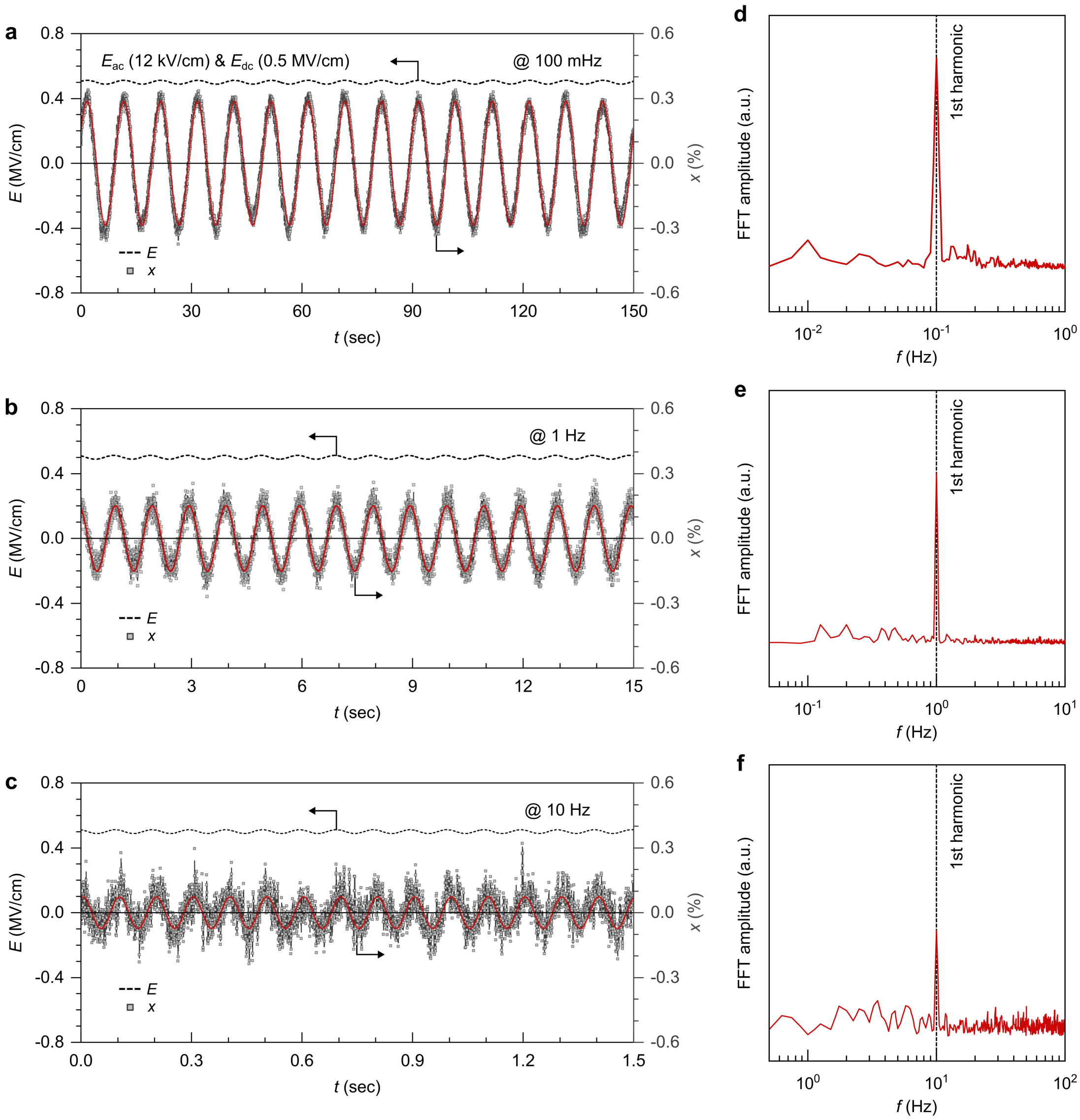

**Fig. S9. Induced piezoelectric effects in the CGO film sample.** a,b,c) The 1st harmonic electromechanical displacements of the CGO film sample, excited by simultaneously applying $E_{AC}$ (= 12 kV/cm) and $E_{DC}$ (= 0.5 MV/cm), measured at different frequencies, $f$ = 100 mHz (a), 1 Hz (b), and 10 Hz (c). d,e,f) The corresponding fast Fourier transform (FFT) magnitude spectra of the output signals as a function of $f$. The first-order harmonic mechanical displacement responses of the sample were determined with respect to the frequency of the excited $E_{AC}$ at $f$ = 100 mHz (d), 1 Hz (e), and 10 Hz (f). The current in (a – c) was centered around zero by removing the background and dc strain

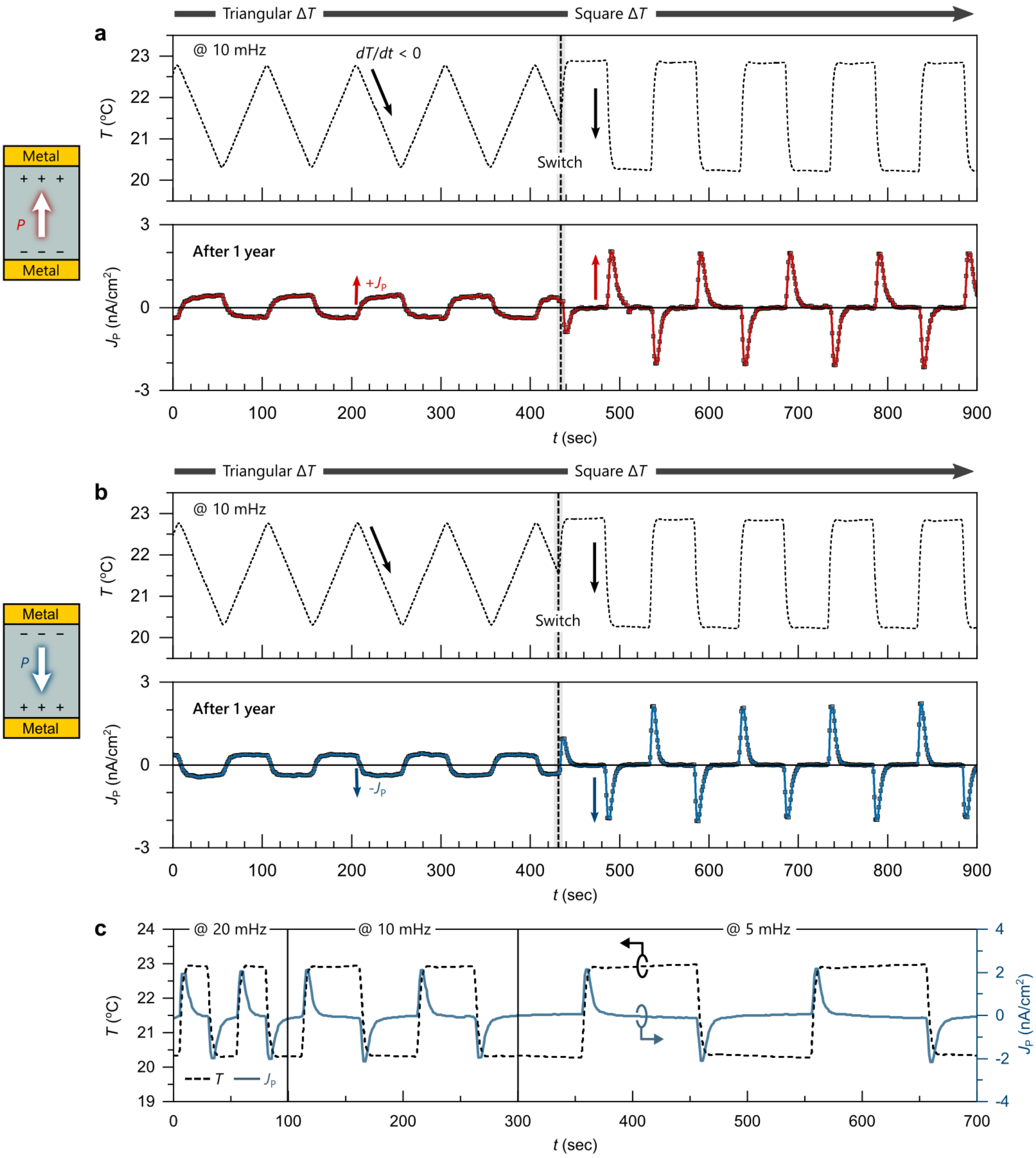

**Figure S10. Persistent pyroelectric effect of the aged, poled CGO film samples with different polarities.** a) Dynamic pyroelectric $J_P$ of a CGO samplewhile continuously applying different temperature waveforms (from triangular to square), at $f = 10$ mHz. The change of the waveform shape over time is denoted by a dashed line. The sample was electrically poled by applying an $E_{DC}$ (-0.8 MV/cm) about one year ago, as schematically illustrated on the left side. b) The same $J_P$ measurement of a CGO sample, electrically poled by applying the opposite $E_{DC}$ (+0.8 MV/cm) about one year ago, as shown on the left side. c, A consistent pyroelectric $J_P$ response of the poled sample, determined by applying continuous square-shaped temperature waveforms at various frequencies (20 mHz – 5 mHz). The current in (a – c) was centered around zero by removing the background current.

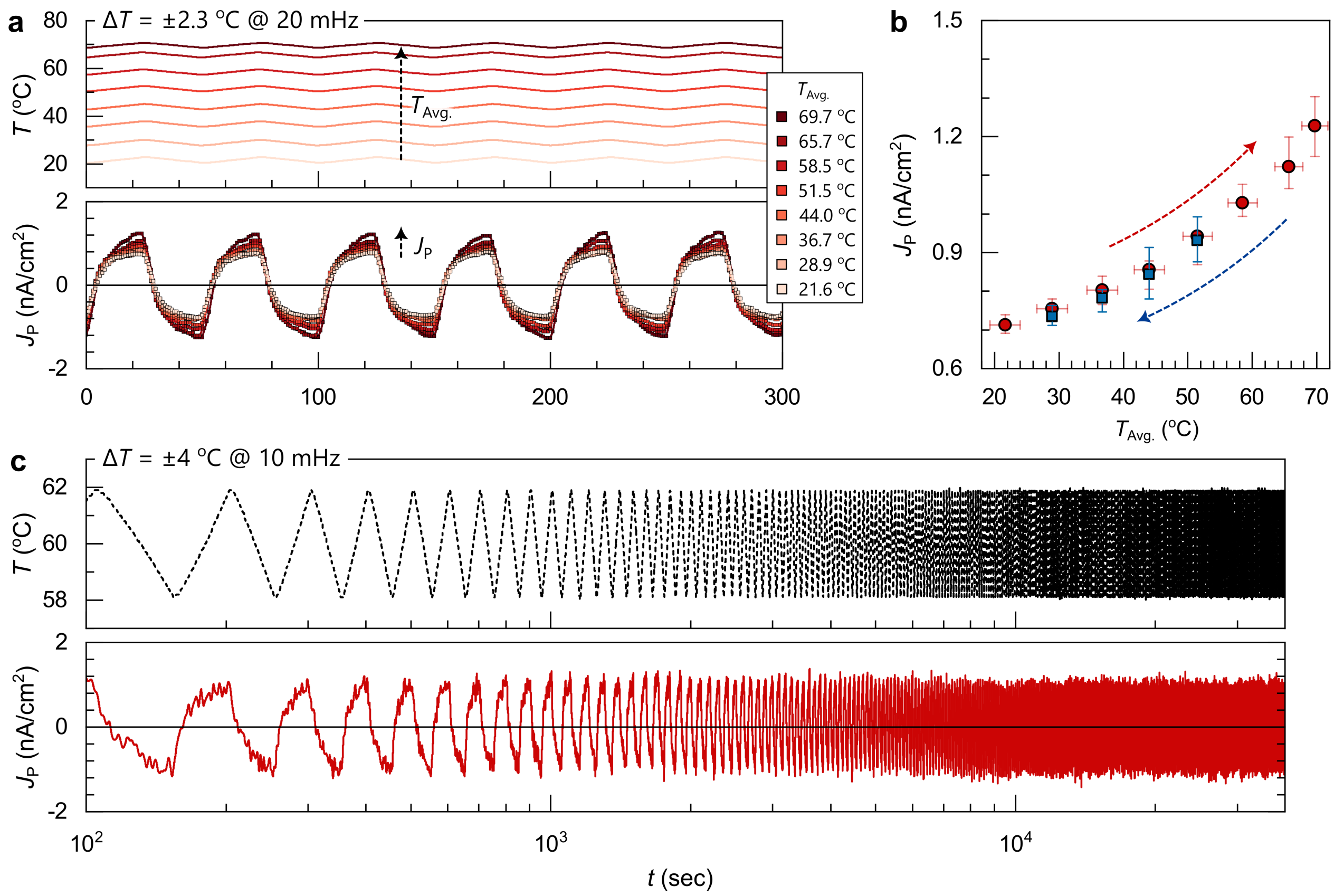

**Figure S11. The pyroelectric current density measurements of the poled CGO samples at elevated temperatures.** a) Pyroelectric current density ($J_p$) measurements of an electrically poled CGO film sample (down-polarized) at elevated average temperatures ($T_{avg.}$) from 21.0 °C to 69.7 °C, recorded under a continuous triangular-shaped temperature modulation of $\Delta T$ =±2.3 °C at 20 mHz. **b)** A plot for the $J_p$ versus $T_{avg.}$ for the CGO sample. **c)** A long-time pyroelectric current measurement of a poled CGO film sample under continuous triangular-shaped temperature waveform application of heating-and-cooling cycles with ΔT = ±4 °C at 10 mHz for more than 10 hours.